\RequirePackage[]{graphicx} 
\graphicspath{{./Figures/}{./Graphs/}{./Drawings/}}

\documentclass[]{arxsub}

\usepackage[utf8]{inputenc}
\usepackage{graphicx}
\usepackage{environ}
\usepackage{listings}
\usepackage{booktabs}
\usepackage{blindtext}
\usepackage{url}
\usepackage{siunitx}

\usepackage[english]{babel}
\usepackage{csquotes}
\usepackage[
  backend=biber,
  natbib = true,
  style=ext-numeric-comp,
  sorting=none,
  minbibnames=3, maxbibnames=6,
  uniquename=false, uniquelist=false,
  giveninits=true, terseinits,
  articlein=false, innamebeforetitle=true,
  isbn=false,
  urldate=long, dateabbrev=false,
  autocite=superscript,
]{biblatex}

\DeclareNameAlias{sortname}{family-given}

\DeclareNameAlias{author}{sortname}
\DeclareNameAlias{editor}{sortname}
\DeclareNameAlias{translator}{sortname}

\DeclareDelimAlias{finalnamedelim}{multinamedelim}

\DeclareFieldFormat
  [article,inbook,incollection,inproceedings,patent,thesis,unpublished]
  {title}{#1}

\renewbibmacro*{journal+issuetitle}{%
  \usebibmacro{journal}%
  \setunit*{\addspace}%
  \iffieldundef{series}
    {}
    {\newunit
     \printfield{series}%
     \setunit{\addspace}}%
  \usebibmacro{issue+date}%
  \setunit{\addsemicolon\addspace}%
  \usebibmacro{volume+number+eid}%
  \setunit{\addcolon\space}%
  \usebibmacro{issue}}

\DeclareFieldFormat[article,periodical]{number}{\mkbibparens{#1}}

\renewbibmacro*{issue+date}{%
  \usebibmacro{date}}

\renewbibmacro*{pubinstorg+location+date}[1]{%
  \printlist{location}%
  \iflistundef{#1}
    {\setunit*{\addsemicolon\space}}
    {\setunit*{\addcolon\space}}%
  \printlist{#1}%
  \setunit*{\addsemicolon\space}%
  \usebibmacro{date}%
  \newunit}

\DeclareFieldFormat{pages}{#1}

\DeclareFieldFormat{doi}{%
  doi\addcolon\space
  \ifhyperref
    {\href{https://doi.org/#1}{\nolinkurl{#1}}}
    {\nolinkurl{#1}}}

\AtEveryBibitem{%
\clearfield{number}%
\clearfield{month}%
\clearfield{day}%
\clearfield{urlyear}%
\clearfield{urlmonth}%
\clearfield{issn}%
\clearfield{series}%
\clearfield{pagetotal}%
\clearfield{eprintclass}%
\ifentrytype{article}{
	\iffieldundef{volume}{}{\clearfield{doi}}%
	\clearfield{eprint}%
	\clearfield{url}%
}{}%
\ifentrytype{book}{%
	\clearfield{eprint}%
	\clearfield{doi}%
	\clearfield{url}%
}{
	\clearname{editor}%
}
\ifentrytype{inproceedings}{%
	\clearfield{eprint}%
	\clearfield{url}%
	\clearfield{doi}%
	\clearfield{publisher}%
	\iffieldundef{booktitle}{}{\clearfield{eventtitle}}%
}{}%
\iffieldundef{eprint}{}{\clearfield{url}}
\iffieldundef{doi}{}{\clearfield{url}}
}

\DeclareMathOperator*{\argmin}{arg\,min}

\newcommand{\figref}[2][]{\hyperref[#2]{Figure~\ref*{#2}#1}}
\newcommand{\tblref}[2][]{\hyperref[#2]{Table~\ref*{#2}#1}}
\renewcommand{\eqref}[1]{\hyperref[#1]{Eq.~\ref*{#1}}}

\newcommand{\bvec}[1]{\mathbf{#1}}

\newcommand{\norm}[1]{\left\lVert#1\right\rVert}

\providecommand{\doi}{}
\renewcommand{\doi}[1]{\href{https://doi.org/#1}{#1}}

\newcommand{\preSubmission}{}

\providecommand{\del}[2][]{}
\providecommand{\dels}[2][]{}
\providecommand{\printfunding}{}

\title{Phase-Pole-Free Images and Smooth Coil Sensitivity Maps by Regularized Nonlinear Inversion}

\corres{Moritz Blumenthal, Graz University of Technology,
Institute of Biomedical Imaging, Stremayrgasse~16/3, 8010~Graz, AUSTRIA, \email{blumenthal@tugraz.at}}

\author[1]{Moritz Blumenthal}{}
\author[1,2,3,4]{Martin Uecker}{}
\authormark{M. Blumenthal and M. Uecker}

\address[1]{Institute of Biomedical Imaging, Graz University of Technology, Graz, Austria}
\address[2]{Institute for Diagnostic and Interventional Radiology, University Medical Center Göttingen, Göttingen, Germany}
\address[3]{DZHK (German Centre for Cardiovascular Research), partner site Lower Saxony}
\address[4]{BioTechMed-Graz, Graz, Austria}

\keywords{MRI, image reconstruction, non-linear inverse problems, parallel imaging, phase singularity}

\articletype{Technical Note}

\finfo{
Funded in part by NIH under grant U24EB029240, and by the Deutsche Forschungsgemeinschaft (DFG, German
Research Foundation) - Project-ID 432680300 - SFB 1456.  This work was supported by DZHK
(German Centre for Cardiovascular Research) funding code: 81Z0300115. The research was funded
in whole or in part by the Austrian Science Fund (FWF) 10.55776/F100800.
}

\abstract{
\section{Purpose}
Phase singularities are a common problem in image reconstruction with auto-calibrated
sensitivities due to an inherent ambiguity of the estimation problem.
The purpose of this work is to develop a method for detecting and correcting phase poles
in non-linear inverse (NLINV) reconstruction of MR images and coil sensitivity maps.
\section{Methods}
Phase poles are detected in individual coil sensitivity maps by computing the curl in each pixel.
A weighted average of the curl in each coil is computed to detect phase poles.
Phase pole detection and correction is then integrated into the iteratively regularized Gauss-Newton
method of the NLINV algorithm, which then avoid phase singularities in the reconstructed images.
The method is evaluated for reconstruction of accelerated Cartesian MPRAGE data of the brain
and interactive radial real-time MRI of the human heart.
\section{Results}
Phase poles are reliably removed in NLINV reconstructions for both applications.
NLINV with phase pole correction can reliably and efficiently estimate coil sensitivity profiles
free from singularities even from very small ($7\times7$) auto-calibration (AC) regions.
\section{Conclusion}
NLINV emerges as an efficient and reliable tool for image reconstruction and coil
sensitivity estimation in challenging MRI applications.
}

\begin{document}

\maketitle

\section{Introduction}
\label{sec:introduction}

MR images are inherently complex-valued. While most clinical applications use
the magnitude images for diagnosis, the phase carries information about the
local magnetic field experienced by the spins during the acquisition.  This
includes contributions from local tissue susceptibility, flow, and chemical
shift. Extracting this information forms the basis of many advanced
MRI applications.

In practice, the MR signal is usually measured by phased array coils
\cite{Roemer_Magn.Reson.Med._1990}, such that the complex-valued image is
multiplied with the complex-valued coil sensitivity profiles. To recover the
original image, the coil sensitivity profiles need to be known absolutely.
However, from the measured data itself, only relative coil sensitivity profiles
can be estimated \cite{Walsh_Magn.Reson.Med._2000,Ying_Magn.Reson.Med._2007,Uecker_Magn.Reson.Med._2008,Uecker_Magn.Reson.Med._2014,Lobos_IEEETrans.Med.Imaging_2024}.
When absolute coil sensitivity profiles are not available, a channel with
relatively homogeneous sensitivity can arbitrarily be chosen as a reference, or
coils can be normalized to have a root-sum-of-squares (RSS) of one such that coil
combination results in a complex-valued image with a magnitude corresponding to
the conventional RSS coil combination \cite{Roemer_Magn.Reson.Med._1990}.

The latter does not define a phase. Normalization techniques to select a phase
include
selection of an arbitrary coil as reference \cite{Griswold_Proc.Annu.Meet.ISMRM_2002,Uecker_Magn.Reson.Med._2014,Lobos_IEEETrans.Med.Imaging_2024}, 
the construction of virtual reference coils
\cite{Buehrer_Magn.Reson.Med._2007,Buehrer_Proc.Annu.Meet.ISMRM_2009,Parker_Magn.Reson.Med._2014,Bilgic__2016}
or normalization such that the image is roughly real-valued
by shifting the phase into the coil sensitivities \cite{Inati_Proc.Annu.Meet.ISMRM_2013,Uecker_Magn.Reson.Med._2017}.
While the latter complicates
processing for applications which require the image phase,
the former effectively sets the image phase to the phase of the respective coil image.
If this coil image contains phase poles, this leads to phase poles, i.e. singularities, in
the images (c.f. Supporting Figure \ref{supfig:espirit}), which are also called open-ended fringe lines in the
phase-unwrapping literature
\cite{Chavez_IEEETrans.Med.Imaging_2002,Robinson_NMRBiomed._2017,Committee_Magn.Reson.Med._2024}
and could be misinterpreted as microhemorrhages in susceptibility weighted imaging
\cite{Li_J.Magn.Reson.Imaging_2015}.

In the SENSE
\cite{Pruessmann_Magn.Reson.Med._1999,Pruessmann_Magn.Reson.Med._2001}
formulation of parallel imaging, coil sensitivity profiles are used to unfold
aliased images.  Nowadays, SENSE reconstruction is often formulated as a linear
inverse problem.  In case of $\ell_2$-regularization,
the reconstruction is invariant to the phase of the images. However, if SENSE is combined with
compressed sensing \cite{Block_Magn.Reson.Med._2007}, the reconstruction may be
sensitive to the phase of the images such as for $\ell_1$-Wavelet
regularization \cite{Lustig_Magn.Reson.Med._2007}.
To overcome phase problems in compressed sensing reconstructions, regularization terms that
are explicitly designed to be phase-invariant have been proposed.
\cite{FengZhao_IEEETrans.Med.Imaging_2012,Ong_Magn.Reson.Med._2017,Liu_IEEETrans.Comput.Imaging_2021}
k-Space-based parallel imaging reconstructions
\cite{Griswold_Magn.Reson.Med._2002,Lustig_Magn.Reson.Med._2010,Haldar_IEEETrans.Med.Imaging_2014,Shin_Magn.Reson.Med._2014}
are invariant to the image phase, as selection of the image phase is postponed to the coil combination step, but complicate the design
of explicit image-based regularization terms.

NLINV \cite{Uecker_Magn.Reson.Med._2010} is a parallel imaging reconstruction
method which jointly estimates the image and the coil sensitivity profiles by
regularized non-linear inversion.  NLINV penalizes the Sobolev norm of the coil
sensitivities, which leads to smooth coil sensitivities.

Due to this smoothness, coil sensitivities can be efficiently estimated on a low resolution grid and
interpolated to target resolution \cite{Holme_Proc.Annu.Meet.ISMRM_2023} without the danger
of introducing artifacts due to non-smooth phase variations. As NLINV can directly work with
non-Cartesian data, it can be used for coil sensitivity estimation in this setting, and is especially
useful in 3D where computational cost is an issue
\cite{Maier_Magn.Reson.Med._2018,Zeng_SignalProcessing:ImageCommunication_2020,Dubljevic_Magn.Reson.Med._2024,Lee_Magn.Reson.Med._2024,Tasdelen_Magn.Reson.Med._2024,Klimes_Magn.Reson.Med._2025,Feizollah_Magn.Reson.Med._2025,Vu_Magn.Reson.Med._2025,Vu_Magn.Reson.Med._}.
Various extensions to NLINV have been developed \cite{Knoll_Magn.Reson.Med._2012,Rosenzweig_Magn.Reson.Med._2018,Holme_Sci.Rep._2019},
including for real-time MRI \cite{Uecker_NMRBiomed._2010,Uecker_Magn.Reson.Med._2010} and
deep-learning-based reconstruction \cite{Leynes_IEEETrans.Med.Imaging_2024,Blumenthal_Magn.Reson.Med._2024},
it has been adapted to various specific applications
\cite{Joseph_NMRBiomed._2011,Niebergall_Magn.Reson.Med._2013,Xu_Magn.Reson.Med._2013,Joseph_J.Magn.Reson.Imaging_2014,
Untenberger_Magn.Reson.Med._2016,Wang_TechnolCancerResTreat_2017,Chen_ApplMagnReson_2017,Dong_NMRBiomed._2023,Xiang_J.Magn.Reson.Imaging_2023,Wang_Magn.Reson.Med._2025,Brisson_ZeitschriftfurMedizinischePhysik_2022},
and used in clinical research
\cite{Zhang_J.Magn.Reson.Imaging_2012,Bassett_NMRBiomed._2014,Olthoff_GastroenterologyResearchandPractice_2014,Carstens_Neurology_2015,Liu_BMEFront._2021,Isaieva_SciData_2021,Laubrock_EuropeanJournalofRadiologyOpen_2022,Liu_BMEFront._2021}.
Despite its successful application in many challenging scenarios, NLINV still
suffers from one problem that prevents its use in routine applications.
 NLINV
solves a non-linear optimization using an iterative method which can get
trapped in local minima. This problem seems directly related to the phase
ambiguity described above, as the local minima observed in practice always
contain phase poles in the image and conjugate phase poles in the coil sensitivity
profiles. This then leads to black holes in regularized reconstructions
\cite{Wang_Magn.Reson.Med._2018,Holme_Sci.Rep._2019,Voit_Quant.ImagingMed.Surg._2019}.
If a body coil as reference is available, the non-linear
inversion can be initialized to deliver phase-pole-free images
\cite{Voit_Quant.ImagingMed.Surg._2019}, but a body coil reference is
not always available. Extending NLINV to ENLIVE \cite{Holme_Sci.Rep._2019} provides
artifact free magnitude images by simultaneously reconstructing multiple images
corresponding to multiple sets of coil sensitivities similar to ESPIRiT
\cite{Uecker_Magn.Reson.Med._2014}. However, the image phase may still contain
phase poles.

In this work, we present a method to detect convergence of NLINV to a local
minimum with phase poles by detecting phase poles and correcting
them by a global optimization step, similar to a re-initialization of the algorithm.
In this global optimization step the phase
pole in the image is removed by multiplying a conjugate pole to the image
while multiplying the pole itself on the coil sensitivity profiles.
While the method is not restricted to NLINV, it is especially well-suited for NLINV since slightly misaligned corrections will be refined
in consecutive iterations.
We evaluate NLINV with phase pole correction on Cartesian brain data and on radial, interactive
real-time MRI data.  Moreover, we show that NLINV can reliably and
efficiently estimate coil sensitivity profiles free from singularities
from very small ($7\times7$) auto-calibration (AC) regions.


\section{Methods}
\label{sec:methods}

The MR signal $\bvec{s}(\bvec{k})\in \mathbb{C}^{N_C}$ measured by a
phased-array coil with $N_C$ coil elements is the Fourier transform of the
product of the coil sensitivity $\bvec{c}(\bvec{r})\in \mathbb{C}^{N_C}$ and
the image $\rho(\bvec{r})\in \mathbb{C}$ at position $\bvec{r}\in
\mathbb{R}^d$, i.e.
\begin{align}
	\bvec{s}(\bvec{k}) = \int_{\mathbb{R}^d} \mathrm{d}^d\bvec{r} \, \bvec{c}(\bvec{r}) \rho(\bvec{r}) e^{-i2\pi \bvec{k} \cdot \bvec{r}}\,.
\end{align}
Here, phase due to the transmit coil is absorbed into $\bvec{c}$
such that the phase $\phi(\bvec{r})$ of the image $\rho(\bvec{r})$
is the time-integrated local Larmor frequency, i.e. $\phi(\bvec{r}) = \int_0^{t_\mathrm{E}} \omega_L(\bvec{r}, t) \mathrm{d}t$.  As only the
product of image and coil sensitivity is measured, both cannot be uniquely
determined from the data but only up to a complex scaling function
$\vartheta(\bvec{r})\in \mathbb{C}\setminus\{0\}$, i.e. if $\rho$ and
$\bvec{c}$ are consistent with the data, then $\rho'=\vartheta^{-1}\rho$ and
$\bvec{c}'=\vartheta\bvec{c}$ are consistent, too.  To break this ambiguity, a
normalization of the coil sensitivities is required.

\subsection{Phase Poles and Singularities in MRI}

\begin{figure}
\includegraphics[width=\linewidth]{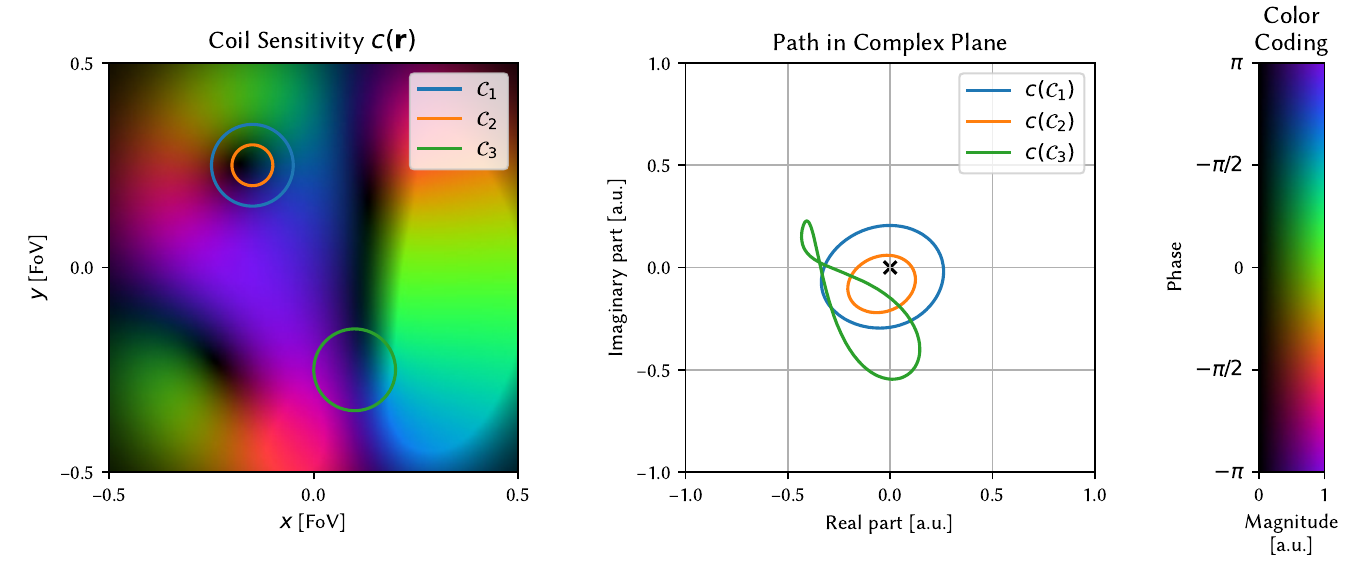}
\caption{Example for a phase singularity in the coil sensitivity profile. The
curves $\mathcal{C}_1$ and $\mathcal{C}_2$ enclose the same phase singularity
such that their image $c(\mathcal{C}_1)$ and $c(\mathcal{C}_2)$ wind once
around the origin in the complex plane. The curve $\mathcal{C}_3$ does not
enclose the singularity and therefore has a winding number of zero.}
\label{fig:phase_singularity} \end{figure}

Let $\Psi(\bvec{r})=A(\bvec{r})e^{-i\phi(\bvec{r})}$ be a complex field, such as an image or coil sensitivities, with
amplitude $A(\bvec{r})$ and phase $\phi(\bvec{r})$.  A phase pole (or phase
singularity) is a point $\bvec{r}_0$ with undefined phase $\phi(\bvec{r})$
further characterized by the condition that a closed loop curve integral of
the phase gradient $\nabla \phi$ along a curve $\mathcal{C}\in\mathbb{R}^d$
around the point is not zero, i.e.
\begin{align}
	0 \neq S_\mathcal{C}=\frac{1}{2\pi}\oint_{\mathcal{C}} \nabla \phi(\bvec{s}) \mathrm{d}\bvec{s}~. \label{eq:phase_singularity}
\end{align}
By interpreting the complex number $\Psi(\bvec{r})$ as a point in the complex
plane, the integral $S_\mathcal{C}$ corresponds to the winding number of the
curve $\Psi(\mathcal{C})\in\mathbb{C}$
\cite{Berry_LesHouchesLectureSeriesSessionXXXV_1981}.
For a smooth field $\Psi(\bvec{r})$, a phase pole in $\bvec{r}_0$ implies a vanishing
amplitude $A(\bvec{r}_0)=0$.  An example for a
phase pole in a coil sensitivity profile and lifting a curve $\mathcal{C}$ to
the complex plane is shown in \figref{fig:phase_singularity}.

It should be noted that the phase of the image $\rho(\bvec{r})$ originates from
the time-integrated local Larmor frequency, i.e. $\phi(\bvec{r}) =
\int_0^{t_\mathrm{E}} \omega_L(\bvec{r}, t) \mathrm{d}t$ and is therefore
well-defined and single-valued in regions with non-vanishing signal. Hence, the
true, i.e. noise-free and non-discretized, image should
not have phase singularities.  In contrast, the phase of the coil sensitivities
$c(\bvec{r})$ can have phase singularities \cite{Inati_Proc.Annu.Meet.ISMRM_2013}.
Recalling the principle of reciprocity, the complex valued coil sensitivity
$c(\bvec{r})$ corresponds to the transversal components of the magnetic field
$\mathcal{B}(\bvec{r})$ generated by the coil, i.e. $c(\bvec{r}) \propto
\mathcal{B}_{\mathrm{x}} - i \mathcal{B}_{\mathrm{y}}$
\cite{Hoult_J.Magn.Reson._1976a}. An example for a phase singularity is the
field of a magnetic dipole $\bvec{m}=m\hat{e}_z$ in the origin. For $z\neq 0$,
the coil sensitivity profile in polar coordinates is give by $c(r,\phi, z) =
\lvert c(r,z) \rvert e^{i\phi}$. It should be noted that even if the physical
coil sensitivities have no phase singularities, coil compression
\cite{Buehrer_Magn.Reson.Med._2007,Zhang_Magn.Reson.Med._2013,Daeun_Magn.Reson.Med._2021}
techniques can introduce phase singularities in virtual coils (c.f. Supporting Figure \ref{supfig:espirit}).

\subsection{NLINV Reconstruction}

NLINV \cite{Uecker_Magn.Reson.Med._2008} is a parallel imaging reconstruction
technique which jointly estimates the image $\rho$ and the coil sensitivity
maps $\bvec{c}$.  It formulates the reconstruction as a non-linear inverse
problem with the forward model $F(\rho, \bvec{c}) = \mathcal{PF}\left(\bvec{c}
\odot \rho\right)$, where $\odot$ is the Hadamard product, i.e. point wise multiplication,
$\mathcal{F}$ is the Fourier transform and
$\mathcal{P}$ is the projection to the Cartesian or non-Cartesian sampling
pattern.
The regularized reconstruction can then be formulated as the non-linear
optimization problem
\begin{align}
	(\rho, \bvec{c}) = \argmin_{\rho, \bvec{c}} \lVert \bvec{y} - F(\rho, \bvec{c})\rVert_2^2 + \alpha \lVert \rho \rVert_2^2 + \alpha \lVert W \bvec{c}\rVert_2^2\,, \label{eq:objective_nlinv}
\end{align}
where the $\ell_2$-norm of the image and the Sobolev norm of the coil sensitivities
are penalized. Here, the Sobolev norm is represented as a weighted $\ell_2$-norm with a
weighting matrix $W$ that penalizes the high frequency components of the coil
sensitivities.  For efficient computation, the optimization is preconditioned
by introducing
\begin{align}
	\tilde{\bvec{c}} &= W^{-1} \bvec{c} &\text{and} &&
	\tilde{F}(\rho, \tilde{\bvec{c}}) = F(\rho, W \tilde{\bvec{c}})\,,
\end{align}
such that \eqref{eq:objective_nlinv} is equivalent to
\begin{align}
	(\rho, \tilde{\bvec{c}}) = \argmin_{\rho, \tilde{\bvec{c}}} \lVert \bvec{y} - \tilde{F}(\rho, \tilde{\bvec{c}} )\rVert_2^2 + \alpha \lVert \rho \rVert_2^2 + \alpha \lVert \tilde{\bvec{c}}\rVert_2^2\,. \label{eq:objective_nlinv_precond}
\end{align}
This optimization problem is solved using the iteratively regularized
Gauss-Newton method (IRGNM), where the regularization parameter $\alpha$ is
chosen for every iteration $k$ such that $\alpha^{(k)}$ exponentially decays
from one to $\alpha_{\mathrm{min}}$.
In each Gauss-Newton step, the non-linear
forward model $\tilde{F}$ is linearized around the current estimate $(\rho^{(k)},
\tilde{\bvec{c}}^{(k)})$. In case of $\ell_2$-regularization, the inner problem
is quadratic problem and can be solved using a conjugate gradient method
with the corresponding regularization parameter $\alpha^{(k)}$.

To favor real-valued images, NLINV is initialized by setting the image to constant one and the coil sensitivities to zero.
The k-space data is normalized such that $\norm{\bvec{y}}_2 = 100$ which is empirically found to yield good results for 2D imaging.
The final reconstruction is usually rescaled to yield the typical RSS scaling, i.e.
\begin{align}
	\rho &\leftarrow \rho \cdot \sqrt{\sum_{i=1}^{N_C} |\bvec{c}_i|^2}
	&\text{and} &&
	\bvec{c} &\leftarrow \bvec{c} \cdot \frac{1}{\sqrt{\sum_{i=1}^{N_C} |\bvec{c}_i|^2}}.
	\label{eq:normalize_rss}
\end{align}

Since the objective function in \eqref{eq:objective_nlinv_precond} is non-convex,
the Gauss-Newton method can converge to local minima.
One class
of local minima are solutions with a phase singularity in the image $\rho(\bvec{r})$.
Even though these phase singularities are not physically meaningful, they can occur
when they are compensated by a conjugate singularity in the coil sensitivity
due to the inherent ambiguity in the problem.
Due to the smoothness penalty of the coil sensitivities, the magnitude of the
coil sensitivity must be zero at the phase singularity.
In a reconstruction with regularization, this leads
to a black hole, i.e. vanishing signal in the
reconstructed image \cite{Wang_Magn.Reson.Med._2018,Holme_Sci.Rep._2019,Voit_Quant.ImagingMed.Surg._2019}.

\subsection{Phase-Pole Correction in NLINV}

To avoid phase pole, we propose to detect and remove artificial poles
during the iterative reconstruction.  If a pole is present at location $\bvec{r}_0$
in the image, there will be conjugate poles in the coil sensitivity
profiles.  The pole can then be removed by multiplying the
image with the inverse pole and the coil sensitivities with the pole. For this,
we use the phase vortex $\vartheta^\pm_{\bvec{r}_0}(\bvec{r})$, defined by
\begin{align}
	\vartheta^\pm_{\bvec{r}_0}(\bvec{r}) = \frac{(\bvec{r} - \bvec{r}_0)_x \pm i (\bvec{r} - \bvec{r}_0)_y}{\lVert \bvec{r} - \bvec{r_0} \rVert}=e^{\pm i\phi(\bvec{r}-\bvec{r}_0)}\,, \label{eq:phase_vortex}
\end{align}
where the phase $\phi$ corresponds to the phase in polar coordinates with
respect to the point $\bvec{r}_0$.  As NLINV does not work directly with the
coil sensitivities $\bvec{c}$ but with the preconditioned coil sensitivities
$\tilde{\bvec{c}}$, the correction is applied on $\bvec{c}$ and transformed to
$\tilde{\bvec{c}}$ by means of a regularized pseudo inverse $W^+$, i.e.
\begin{align}
	\rho^{(k)} & \leftarrow \vartheta^\pm_{\bvec{r}_0} \odot \rho^{(k)}\,;
	&
	\tilde{\bvec{c}}^{(k)} &\leftarrow W^{+}\left[\vartheta^{\mp}_{\bvec{r}_0} \odot W\tilde{\bvec{c}}^{(k)}\right]\,. \label{eq:phase_pole_correction}
\end{align}
Multiple poles can be corrected simultaneously by using the product of the corresponding vortices.

\begin{figure}
\includegraphics[width=\linewidth]{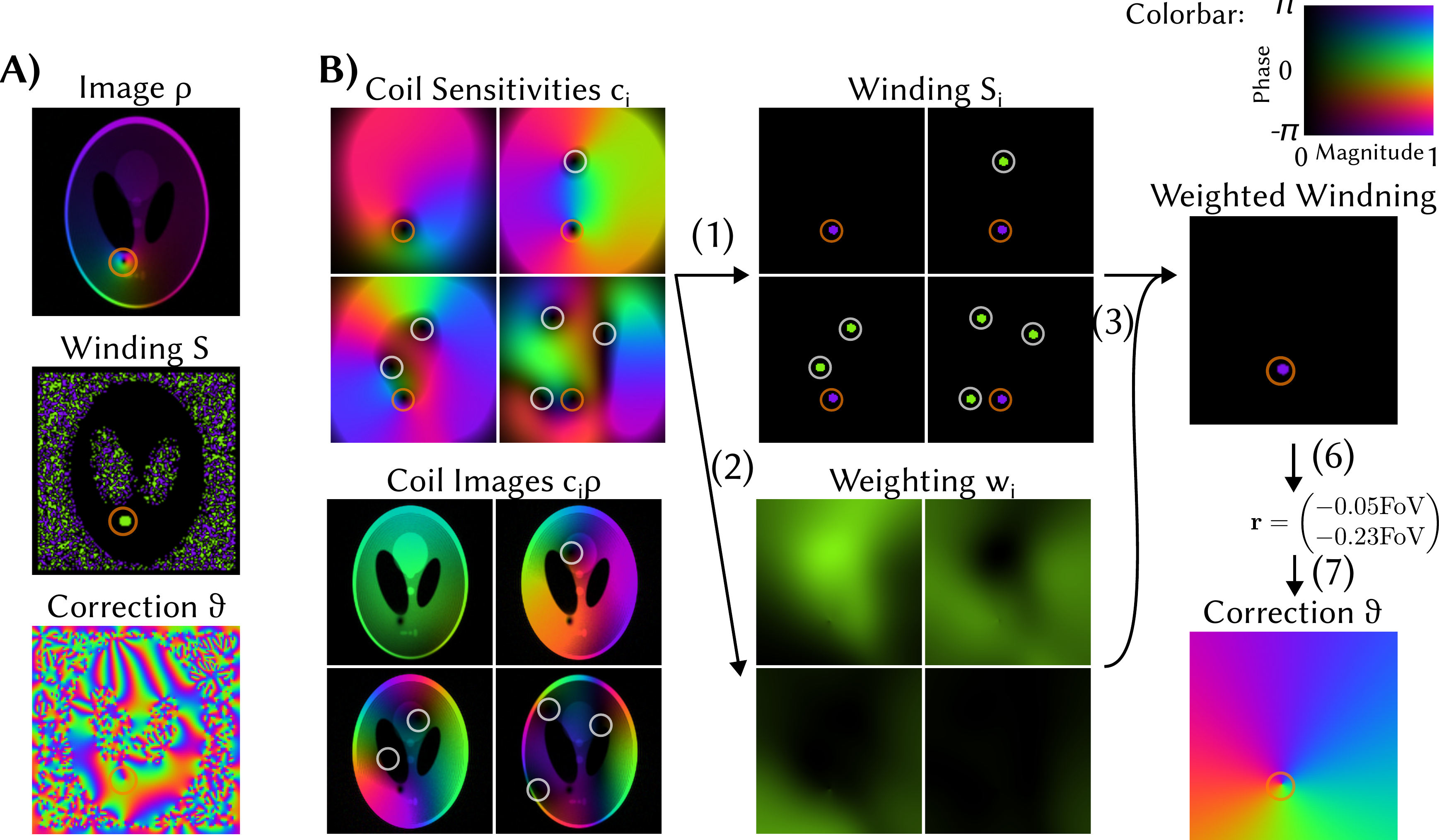}
\caption{Image-based (A) and coil-based-phase (B) phase pole detection. The
image-based detection is not robust against noise leading to a random
correction function. In the coil-based detection,
multiple poles are detected. Those corresponding to true singularities in the
sensitivities are marked by white circles and the corresponding poles are
also contained in the coil-images. After taking the weighted average, only the
false pole which is present in all coils and in the image is detected, leading
to a single factor in the correction function. In this example, we do not show
the thresholding step in (4) as it has no visible effect. Further,
we skip the
morphological closing in step (5), as it would connect all noise poles in the
image-based detection. We refer to Supporting Figure \ref{supfig:detection_real} for an example with real data.}
\label{fig:figure_detection}
\end{figure}

Phase poles can be detected by computing the winding number $S_\mathcal{C}$ in \eqref{eq:phase_singularity}.
The integral can be discretized by splitting the curve $\mathcal{C}$ into $N$ segments $\mathcal{C}_i$ connecting the points $\bvec{r}_i$ and $\bvec{r}_{i+1}$ with $\bvec{r}_{N} = \bvec{r}_0$ on the discrete grid, i.e.
\begin{align}
	S_\mathcal{C}&=\frac{1}{2\pi}\oint_{\mathcal{C}} \nabla \phi(\bvec{s}) \mathrm{d}\bvec{s}= \frac{1}{2\pi}\sum_{i=0}^{N-1} \int_{\mathcal{C}_i} \nabla \phi(\bvec{s}) \mathrm{d}\bvec{s} \notag\\
	&\approx \frac{1}{2\pi} \sum_{i=0}^{N-1} \left( \phi(\bvec{r}_{i+1}) - \phi(\bvec{r}_i) \right)\,.
\end{align}
Even if the true image $\rho(\bvec{r})$ does not have phase
singularities, the discretized image can show phase singularities due to too
coarse discretization or noise \cite{Chavez_IEEETrans.Med.Imaging_2002}.
Because the phase pole detection is not robust for images, we propose to detect
phase poles in the coil sensitivities $\bvec{c}^{(k)}$.  Assuming a pole in the
image, all coil sensitivities should contain the conjugate of the pole as long
as a coil does not happen to have a true phase singularity at the same position.
As the coil sensitivities are smooth, they do not contain noise corrupting the
phase pole detection.
Our detection algorithm is based on the following steps
 which are illustrated in \figref{fig:figure_detection} for a schematic example and in
Supporting Figure \ref{supfig:detection_real} for a real example:
\begin{enumerate}
\item For each coil sensitivity $c_i$, the winding number
$S_{i,\mathcal{C}}(\bvec{r})$ is computed for each pixel $\bvec{r}$
for a curve $\mathcal{C}$ connecting the grid points $\bvec{r}_j\quad (j=0\dots N-1)$ that approximate a small circle around $\bvec{r}$.
The diameter $d$ of the circle must be
chosen sufficiently large such that for slightly misaligned poles the detected
area in different coils overlaps (c.f. Supporting Figure \ref{supfig:detection_real}).
Too large diameters may lead to overlaps of independent phase poles and spoil spatial resolution of the detection.
We empirically choose $d=\SI{0.05}{FoV}$ which worked well for high and low resolution detection.
\item A weighting function $w_i(\bvec{r}) = \lvert c_i(\bvec{r})
\rvert^2$ is computed by taking the magnitude squared of the sensitivity maps.
This function weights down contribution from physical poles in coil sensitivities as those imply vanishing amplitude.
\item The weighted average $S_{\mathcal{C}}(\bvec{r})=\frac{\sum_i w_i(\bvec{r})
S_{i,\mathcal{C}}(\bvec{r})}{\sum_i w_i(\bvec{r})}$
is computed. As visualized in \figref{fig:figure_detection}, this step suppresses
all detected poles that are true poles in the coil sensitivities.
\item The weighted map $S_{\mathcal{C}}(\bvec{r})$ is thresholded to values with magnitude larger than a threshold $t$; we use $t=\frac{1}{2}$.
This yields a binary map indicating that a voxel is close to a pole in the image. Each of the $M$ connected component of this binary map is considered to correspond to a single pole $j=1,\dots, M$.
\item

Separated but close connected components are merged by a morphological closing operation with a disk of diameter $d_{\mathrm{closing}}=d$.
This step prevents over-compensation of phase poles by a double correction if one
pole is incorrectly interpreted as two close-by poles (c.f. Supporting Figure \ref{supfig:detection_real}).
\item
The centers of mass $\bvec{r}_j$ of the remaining connected components are computed, corresponding to the location of the $M$ respective poles.
\item For each detected pole, the corresponding phase vortex
$\vartheta^\pm_{\bvec{r}_j}(\bvec{r})$ is computed. Its sign depends on
the sign of the weighted average. All phase vortices are
multiplied to yield the final correction function
$\vartheta(\bvec{r}) = \prod_{j=1}^M
\vartheta^{\pm_j}_{\bvec{r}_j}(\bvec{r})$.
\item A global phase for the correction function is computed such that the influence of the correction on the image is minimized, i.e.
\begin{align}
	\vartheta(\bvec{r}) &\leftarrow \vartheta(\bvec{r}) \cdot \langle \rho, \vartheta\odot\rho\rangle / \lvert\langle \rho, \vartheta\odot\rho\rangle\rvert\,.
\end{align}
Here, $\langle \cdot, \cdot \rangle$ denotes the inner product.
Intuitively, the inner product averages the complex conjugate of $\vartheta$
weighted by the image power, yielding a final correction that changes the image phase as little as possible in high signal regions.
\end{enumerate}

\subsection{Experiments}

All MRI data have been acquired with written informed consent and with approval
of the local ethics committee on a Magnetom VIDA 3T (Siemens Healthineers).
Phase pole detection and correction was integrated in BART
\cite{Uecker__2013,Uecker__2015} and all reconstructions were performed with
BART.

Fully-sampled 3D Cartesian MPRAGE
($\mathrm{TR}/\mathrm{TE}/\mathrm{TI}=2200/2.46/\SI{900}{ms}$) data of a
healthy volunteer were acquired with a 20 channel head coil. Matrix size was
$256\times256\times208$ and resolution $1\times1\times\SI{0.9}{mm^3}$.  The data
was retrospectively undersampled in the $k_y-k_z$-plane with a $2\times2$
regular pattern and fully-sampled auto-calibration (AC) regions of size
$3\times3$, $7\times7$, and $15\times15$.
The data was pre-whitened \cite{Pruessmann_Magn.Reson.Med._2001}
and coil-compressed to 12 virtual coils.
After an inverse Fourier transform in the readout direction, the data was
reconstructed slice-wise using NLINV with and without phase pole correction.
Twelve Gauss-Newton steps were performed with a minimal regularization parameter
of $\alpha_{\mathrm{min}}=0.001$.
Phase pole detection was performed once after eight
Gauss-Newton steps such that the problem is sufficiently converged and 4 iterations
remain for refinement of the reconstruction.

To evaluate the coil sensitivity profiles estimated by NLINV (with and without
phase pole correction), we compare them with ESPIRiT
\cite{Uecker_Magn.Reson.Med._2014}.  For this, NLINV coil sensitivity profiles
were normalized to have an RSS of one as in ESPIRiT.  Using the fully-sampled
data, a projection test \cite{Uecker_Magn.Reson.Med._2014} was performed.

For this, the coil images $\bvec{m}=\mathcal{F}^{-1}\bvec{y}$ were computed by an inverse Fourier transform.
The coil images were then projected to the subspace spanned by the coil sensitivities by the point-wise projection operator
$P_{\bvec{c}(\bvec{r})}=\norm{\bvec{c}(\bvec{r})}^{-2}\bvec{c}(\bvec{r})\bvec{c}(\bvec{r})^H$.
For well estimated coil sensitivities, the difference to the measured coil images $\bvec{m}$ should be the
whitened measurement noise and not contain any image information.

Moreover, for all undersampling patterns, the resulting coil
sensitivity profiles were used for an $\ell_1$-Wavlet regularized SENSE
reconstruction.  As a reference to evaluate these reconstructions, we used the
fully-sampled data, and coil combined images were computed using ESPIRiT coils
estimated from a $24\times24$ AC region.

To demonstrate the efficiency NLINV as a coil sensitivity estimation method, we
measured the time to estimate the coil sensitivities with a $15\times15$ AC region
for all 256 slices of the dataset using ESPIRiT, PISCO\cite{Lobos_IEEETrans.Med.Imaging_2024} and NLINV with and without
phase pole correction. For all methods, the number of threads used per slice was set to one on an Intel Xeon Gold 6136 CPU.
For PISCO, we used the Matlab implementation\footnote{\url{https://github.com/ralobos/PISCO},
Commit 986a00b} with and without interpolation of the final coil sensitivities from a low resolution grid.
Default parameters were used, except for the threshold of the singular values which,
was set to $0.001$, and the FFT based computation of the calibration matrix, which was disabled
due to a too large approximation error on this small AC region, leading to poor coil sensitivities.
For NLINV, we estimated coils using the full resolution and on a low resolution $48\times48$ grid.

Phase pole correction was also tested for changing coil sensitivities in
interactive real-time MRI.  Radial
FLASH data ($\mathrm{TR}/\mathrm{TE}=2.06/\SI{0.76}{ms}$) was acquired
with a turn based scheme with 13 spokes per frame and 5 turns
\cite{Uecker_NMRBiomed._2010} for in total $\SI{32}{s}$, i.e. 1200 frames.  The
imaging slice was changed interactively during the acquisition to trigger a
phase poles in the reconstruction.  Gradient delays were corrected using RING
\cite{Rosenzweig_Magn.Reson.Med._2018a} and the data was compressed to 8
virtual coils using geometric coil compression
\cite{Zhang_Magn.Reson.Med._2013,Schaten_MagnResonMaterPhy_2024}.  Real-time
images were reconstructed using real-time NLINV
\cite{Uecker_Magn.Reson.Med._2010} implemented in BART's streaming framework
\cite{Schaten_MagnResonMaterPhy_2024} with and without phase pole correction
using 6 Gauss-Newton steps, allowing for reconstruction in real time.
Phase pole correction was performed after the last
step, such that the correction is refined in the following frames.

\section{Results}
\label{sed:results}

\begin{figure}
\includegraphics[width=\linewidth]{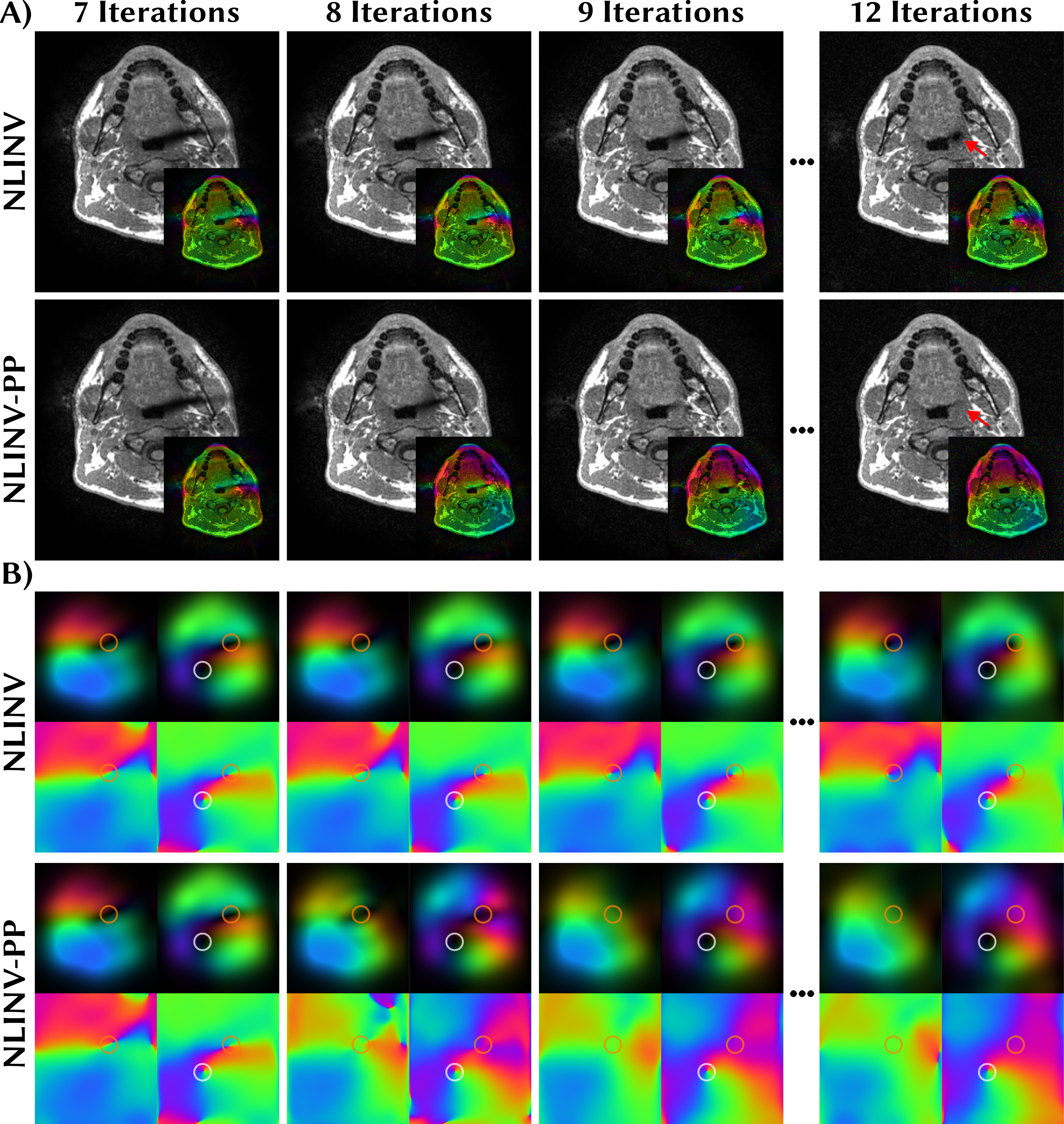}
\caption{NLINV reconstructions (A) and the first two coil sensitivities (B) with and without phase pole correction.
Insets in (A) show the complex image before normalization (\eqref{eq:normalize_rss}).
Coil sensitivities are shown once as complex maps (first row) and once phase only (second row)
to highlight the phase poles which are located in the magnitude regions of the coils.
After eight iterations the phase pole is corrected visible by the changed phase in the image and coils.
After that the magnitude normalizes over the remaining iterations. In the final reconstruction,
the phase pole in the non-corrected reconstruction leads to a black
hole marked by the red arrow.
Orange circles mark the position of the phase pole in the coil sensitivities.
The white circles mark the position of a true phase pole only present in the
2nd coil sensitivity map. Phase is color-coded as in Figure 1.} \label{fig:figure_iterations}
\end{figure}

We present the NLINV reconstructions with and without phase pole correction of
the retrospectively undersampled data in \figref{fig:figure_iterations}.  In the
final reconstruction, the phase pole leads to a black hole as the coils are
vanishing due to the singularity (marked by red arrow).  After eight
iterations, the phase pole is detected and corrected, however, the correction
does not immediately affect the magnitude of the image.  In the phase-pole
corrected reconstruction, the magnitude normalizes after correction over the
remaining four Gauss-Newton steps, leading to a homogeneous image without
singularity.

\begin{figure}
\includegraphics[width=\linewidth]{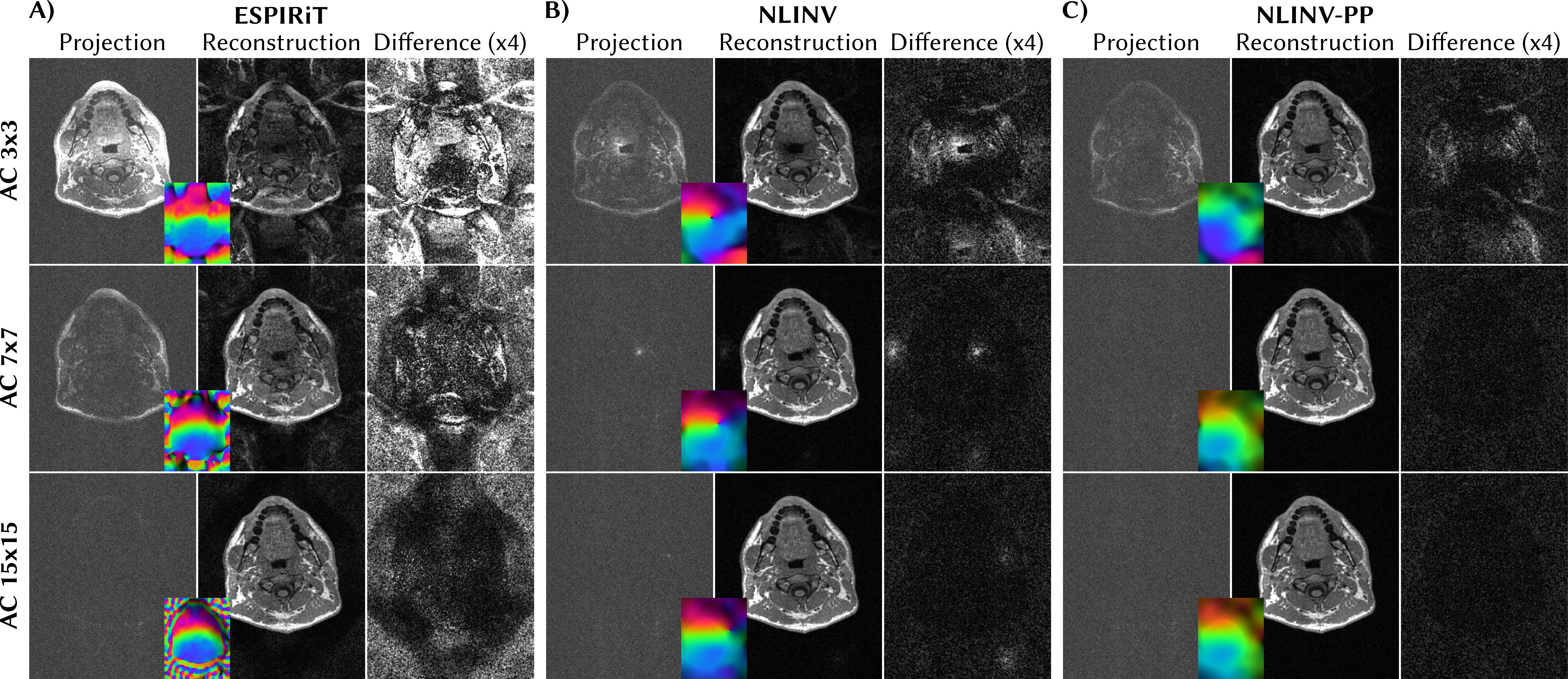}
\caption{Evaluation of coils estimated with ESPIRiT (A), NLINV (B), and NLINV
with phase pole correction (C). The first column shows the
root-sum-of-squares difference between the coil images and their
point-wise projection to the space spanned by the coil sensitivities. The second column shows
reconstructed images using the respective coils in an $\ell_1$-Wavelet
reconstruction. The third column shows the difference of
reconstructions to the fully sampled reference. Rows correspond to
different sizes of full-sampled AC regions.  The inlaid images show
the first estimated virtual coil to visualize the position of the phase pole (phase is color-coded as in Figure 1).  NLINV
provides good coil sensitivities starting from $7\times7$ AC regions, while phase
pole correction removes remaining artifacts in this example. ESPIRiT
requires larger AC regions to achieve similar results.}
\label{fig:figure_projection}
\end{figure}

We present the results of the projection tests, i.e. the root-sum-of-squares difference of the coil images and their projection to the space spanned by the respective coil sensitivities, in the first column of \figref{fig:figure_projection} A), B), and C), respectively. Recall, ideal results only show noise and no image information.

The projection
for NLINV coil sensitivities does not show image structures from AC regions
larger than $7\times7$, while ESPIRiT requires larger AC regions ($15\times15$) to
achieve similar results.  The effect is also visible in the reconstructed
images which show aliasing artifacts for ESPIRiT with small AC-regions.
The phase pole in the coil sensitivities obtained with NLINV without
correction also leads to residual errors which are visible in the projection
and the reconstructed image, and absent for NLINV with correction.

\begin{figure}
\includegraphics[width=\linewidth]{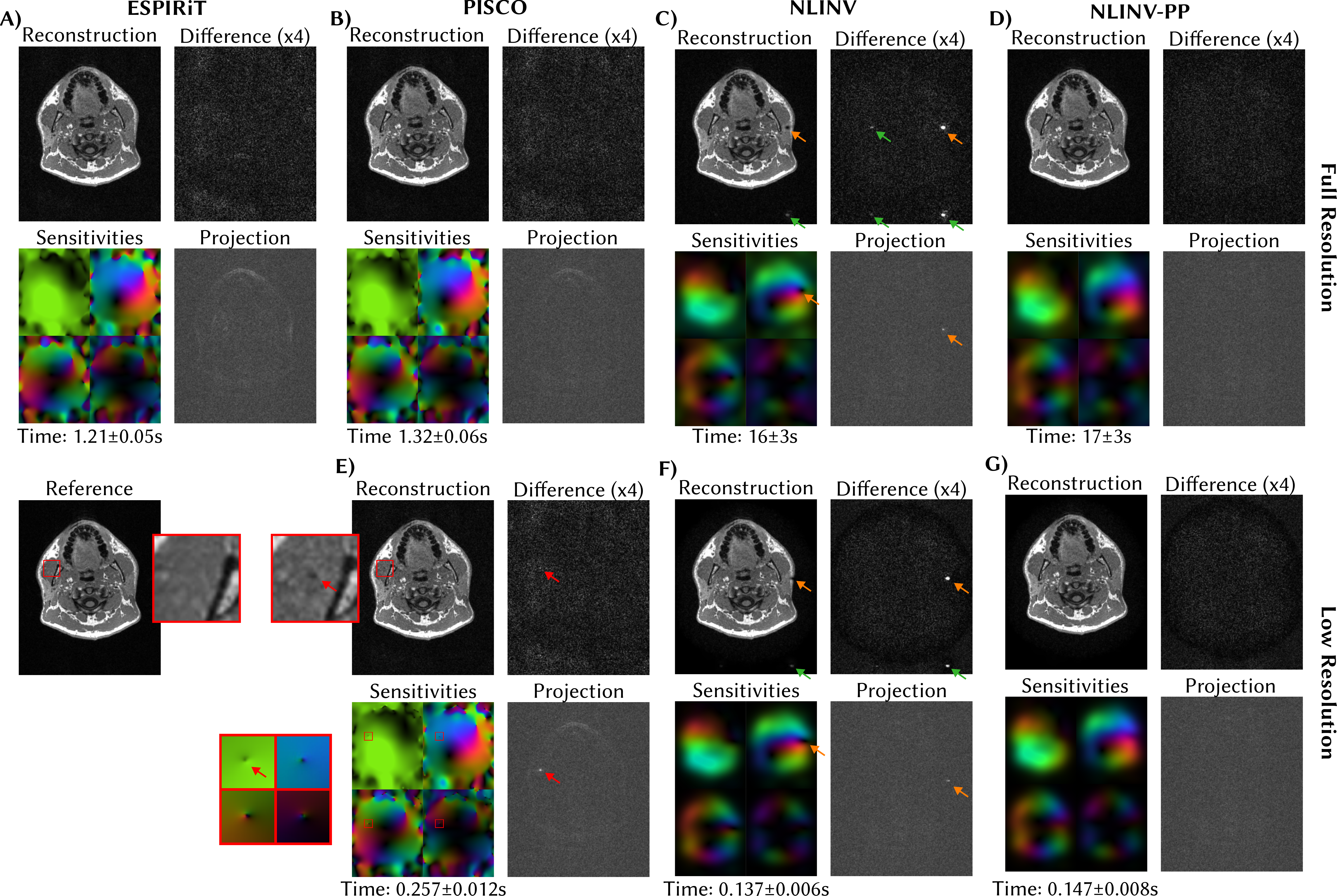}
\caption{
Comparison of different coil sensitivity estimation methods in terms of
estimation time, projection error, and reconstruction error.
In the top row (A-D), coil sensitivities (phase is color-coded as in Figure 1) are estimated on the full resolution grid.
ESPIRiT and PISCO are similarly fast, while NLINV is significantly slower. However, with
phase pole correction, NLINV provides the best coil sensitivities in terms of the
projection test, leading to the lowest difference of the reconstruction to the reference.
Phase poles are marked by orange arrows and corresponding aliasing artifacts
by green arrows. The bottom row (E-G) shows results for coil sensitivities estimated on a
low resolution grid by PISCO and NLINV. Due to interpolation of a non-smooth phase, PISCO
introduces an artifact (red arrow) in the coils, which is visible in the projection test and
leads to an artifact (dark spot visible in the inset) in the reconstruction.
}
\label{fig:figure_speed}
\end{figure}

Results of the measurement of coil sensitivity estimation time, are shown in
\figref{fig:figure_speed}. It can be seen that both, PISCO and NLINV significantly benefit
from use of a low resolution grid for coil sensitivity estimation. For PISCO, the
interpolation of the coil sensitivities to the target resolution introduces an artifact
in the coil sensitivities, probably due to non-smooth phase variations, which is not visible
in the high resolution estimation, but leads to an artifact in the final reconstruction.
We saw similar artifacts in neighboring image slices, probably due to the non-trivial phase
in the jaw region. NLINV intrinsically provides smooth coil sensitivities that can be
interpolated to the target resolution without visually worsening the results of the
projection test compared to the full resolution estimation. The additional overhead
for phase pole correction in NLINV is below 10\% of the total estimation time.

\begin{figure}
\includegraphics[width=\linewidth]{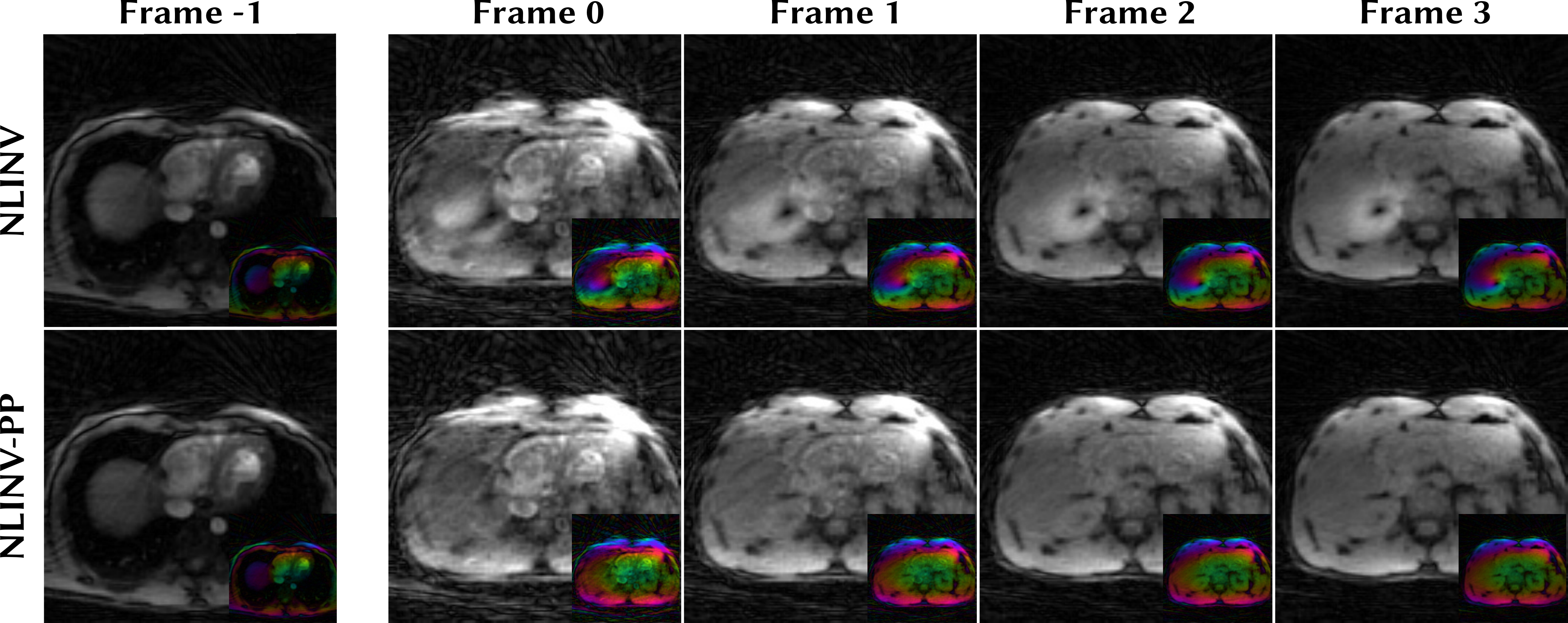}
\caption{Real-time reconstruction with (NLINV-PP) and without (NLINV) phase
pole correction. The imaging slice is interactively changed between
Frame -1 and Frame 0 which in this example causes a phase pole to occur
in the reconstruction. This pole is detected and corrected in the next
frame (Frame 1). Inlays show the complex-valued images with phase
color-coded as in Figure 1.}
\label{fig:figure_realtime}
\end{figure}

Reconstructions of the real-time data are shown in
\figref{fig:figure_realtime}. NLINV with phase pole correction detects and
corrects the phase pole induced by the abrupt change of the imaging slice
between Frame -1 and Frame 0. A movie of the full real-time reconstruction is
provided in the supplementary material.  Reconstruction of the whole time
series takes $\SI{22}{\second}$ without phase pole correction and
$\SI{25}{\second}$ with phase pole correction on an Nvidia H100 GPU.  Both
below the acquisition length of $\SI{32}{\second}$, which is the limit for
real-time reconstruction.

\section{Discussion}
\label{sec:discussion}

In this work, we present a simple method to obtain phase-pole-free images from
NLINV reconstructions. Phase poles are detected by computing the winding
numbers in the coil sensitivities.  If the coils agree on a phase pole at the
same location, it is considered an artificial phase pole.  This technique
is more robust compared to detecting phase poles directly in the image which is
prone to noise in low signal intensity regions.  Nevertheless, the technique
requires three parameters to be set: the diameter $d$ of circles used to
compute the winding number, the threshold $t$ for detecting phase poles, and
the minimum distance $d_{closing}$ between detected phase poles to consider
them as distinct. All our experiments worked well with the same default values
of $d=\SI{0.05}{FoV}$, $t=0.5$, and $d_{\mathrm{closing}}=d$.

The current work is focussed on phase pole correction in NLINV. Principally, the
idea of detecting the phase poles in the coil sensitivities instead of the image
can also be applied to other coil sensitivity estimation methods, e.g., ESPIRiT
(c.f. Supporting Figure \ref{supfig:espirit}). However, the iterative approach of
NLINV is particularly well suited for this task as the coil sensitivities can be
refined after the correction in the following iterations leading to smooth sensitivities.

From a computational perspective, the method is efficient and has negligible, i.e. below 10\%,
overhead compared to the NLINV reconstruction.
The computation of the winding
numbers is trivially vectorized
and can be efficiently performed
on the GPU (if one is available).
This allows us to utilize the method for interactive real-time MRI applications
which are prone to phase poles due to interactive changes of the imaging slice.

A current limitation of the method is its restriction to 2D reconstructions. In
3D, phase poles are not singular points but
form lines around which the phase wraps
\cite{Berry_LesHouchesLectureSeriesSessionXXXV_1981}, also called vortex
lines.

Hence, detection of 2D coordinates of phase poles in 2D turns to
identification of 1D curves in 3D.
If a vortex line is mostly orthogonal to the slices of a 3D volume, its intersection with the slice is a point and the 2D method can be applied slice-by-slice.
However, this will likely fail if the vortex line is approximately parallel to the slices, as the singularity in the plane will be more similar to a phase jump
than to a phase vortex. We plan to extend the method to 3D reconstructions in future work.

Finally, it should be noted that NLINV can also be used to estimate the coil
sensitivity profiles for use in other reconstruction algorithms.  We showed in
this work, that NLINV can estimate high quality coil sensitivity profiles even
from very small ($7\times7$) AC regions.  Compared to ESPIRiT or PISCO, which require
larger AC regions, NLINV has the advantage that it can directly be applied to
non-Cartesian data, that it is computationally cheaperon a low-resolution grid, 
and that it can be more easily accelerated with GPUs \cite{Schaetz_ComputMathMethodM_2017}, motivating
its use in several recent studies \cite{Maier_Magn.Reson.Med._2018,Zeng_SignalProcessing:ImageCommunication_2020,Dubljevic_Magn.Reson.Med._2024,Lee_Magn.Reson.Med._2024,Tasdelen_Magn.Reson.Med._2024,Klimes_Magn.Reson.Med._2025,Feizollah_Magn.Reson.Med._2025,Vu_Magn.Reson.Med._2025,Vu_Magn.Reson.Med._}.
The remaining problem of the algorithm to get stuck in local minima with phase poles
is solved by the proposed phase pole correction method.

\section{Conclusion}

In conclusion, we presented a simple method for detecting
and correcting phase poles that is integrated into iterative
reconstruction with NLINV to simultaneously reconstruct images and
coil sensitivities free of phase singularities. NLINV emerges as an efficient and reliable tool
for image reconstruction and coil sensitivity estimation
in challenging applications.


\section*{Conflict of Interest}
The authors declare no competing interests.

\section*{Data Availability Statement}
In the spirit of reproducible research, the code to reproduce the results of this paper is available at \url{https://gitlab.tugraz.at/ibi/mrirecon/papers/phase-pole} (Version v0.2).
All reconstructions have been performed with BART, available at \url{https://github.com/mrirecon/bart}.
The data used in this study is available at Zenodo \doi{10.5281/zenodo.16737746}.

\section*{Acknowledgements}
We acknowledge initial numerical experiments on phase pole detection by Dr. Tilman J. Sumpf.
\printfunding

\printbibliography

@article{Bassett_NMRBiomed._2014,
  title = {Evaluation of Highly Accelerated Real-Time Cardiac Cine {{MRI}} in Tachycardia},
  author = {Bassett, Elwin C. and Kholmovski, Eugene G. and Wilson, Brent D. and DiBella, Edward V. R. and Dosdall, Derek J. and Ranjan, Ravi and McGann, Christopher J. and Kim, Daniel},
  date = {2014},
  journaltitle = {NMR in Biomedicine},
  volume = {27},
  number = {2},
  pages = {175--182},
  issn = {1099-1492},
  doi = {10.1002/nbm.3049},
  url = {https://onlinelibrary.wiley.com/doi/abs/10.1002/nbm.3049},
  urldate = {2025-11-02},
  langid = {english}
}

@incollection{Berry_LesHouchesLectureSeriesSessionXXXV_1981,
  title = {Singularities in {{Waves}}},
  booktitle = {Les {{Houches Lecture Series Session XXXV}}},
  author = {Berry, Michael},
  editor = {Balian, Roger and Kl\'eman, M and Poirier, J-P},
  date = {1981},
  pages = {453--543},
  publisher = {North-Holland},
  location = {Amsterdam},
  eventtitle = {Ecole d'{{Et\'e}} de {{Physique Th\'eorique}}},
  isbn = {978-0-444-86225-9},
  langid = {english}
}

@inproceedings{Bilgic__2016,
  title = {Block Coil Compression for Virtual Body Coil without Phase Singularities},
  booktitle = {Fourth International Workshop on {{MRI}} Phase Contrast \& Quantitative Susceptibility Mapping},
  author = {Bilgic, B. and Marques, J. P. and Wald, L. L. and Setsompop, K.},
  date = {2016},
  location = {Graz},
  url = {https://www.martinos.org/~berkin/2016_07_30_BCC_abstract.pdf}
}

@article{Block_Magn.Reson.Med._2007,
  title = {Undersampled Radial {{MRI}} with Multiple Coils. {{Iterative}} Image Reconstruction Using a Total Variation Constraint},
  author = {Block, K. T. and Uecker, M. and Frahm, J.},
  date = {2007},
  journaltitle = {Magn. Reson. Med.},
  volume = {57},
  number = {6},
  pages = {1086--1098},
  publisher = {Wiley-Blackwell},
  issn = {1522-2594},
  doi = {10.1002/mrm.21236}
}

@article{Blumenthal_Magn.Reson.Med._2024,
  title = {Self-Supervised Learning for Improved Calibrationless Radial {{MRI}} with {{NLINV-Net}}},
  author = {Blumenthal, Moritz and Fantinato, Chiara and Unterberg-Buchwald, Christina and Haltmeier, Markus and Wang, Xiaoqing and Uecker, Martin},
  date = {2024},
  journaltitle = {Magnetic Resonance in Medicine},
  volume = {92},
  number = {6},
  pages = {2447--2463},
  issn = {1522-2594},
  doi = {10.1002/mrm.30234},
  url = {https://onlinelibrary.wiley.com/doi/abs/10.1002/mrm.30234},
  urldate = {2025-11-02},
  langid = {english}
}

@article{Brisson_ZeitschriftfurMedizinischePhysik_2022,
  title = {A Novel Multipurpose Device for Guided Knee Motion and Loading during Dynamic Magnetic Resonance Imaging},
  author = {Brisson, Nicholas M. and Kr\"amer, Martin and Krahl, Leonie A. N. and Schill, Alexander and Duda, Georg N. and Reichenbach, J\"urgen R.},
  date = {2022-11-01},
  journaltitle = {Zeitschrift f\"ur Medizinische Physik},
  volume = {32},
  number = {4},
  pages = {500--513},
  issn = {0939-3889},
  doi = {10.1016/j.zemedi.2021.12.002},
  url = {https://www.sciencedirect.com/science/article/pii/S093938892100115X},
  urldate = {2025-11-02}
}

@article{Buehrer_Magn.Reson.Med._2007,
  title = {Array Compression for {{MRI}} with Large Coil Arrays},
  author = {Buehrer, Martin and Pruessmann, Klaas P. and Boesiger, Peter and Kozerke, Sebastian},
  date = {2007},
  journaltitle = {Magn. Reson. Med.},
  volume = {57},
  number = {6},
  pages = {1131--1139},
  doi = {10.1002/mrm.21237},
  url = {https://onlinelibrary.wiley.com/doi/abs/10.1002/mrm.21237}
}

@inproceedings{Buehrer_Proc.Annu.Meet.ISMRM_2009,
  title = {Virtual {{Body Coil Calibration}} for {{Phased-Array Imaging}}},
  booktitle = {Proceedings of the {{Annual Meeting}} of {{ISMRM}}},
  author = {Buehrer, M and Boesiger, P and Kozerke, S},
  date = {2009},
  volume = {17},
  location = {Honolulu},
  langid = {english}
}

@article{Carstens_Neurology_2015,
  title = {Real-Time {{MRI}} for Evaluation of Dysphagia in Inclusion Body Myositis ({{IBM}})({{P2}}. 015)},
  author = {Carstens, P.-O. and Zhang, S. and Olthoff, A. and Bremen, E. and Lotz, J. and Frahm, J. and Schmidt, J.},
  date = {2015},
  journaltitle = {Neurology},
  volume = {84},
  pages = {P2. 015},
  publisher = {AAN Enterprises},
  issue = {14 Supplement}
}

@article{Chavez_IEEETrans.Med.Imaging_2002,
  title = {Understanding Phase Maps in {{MRI}}: A New Cutline Phase Unwrapping Method},
  shorttitle = {Understanding Phase Maps in {{MRI}}},
  author = {Chavez, S. and Xiang, Qing-San and An, L.},
  date = {2002-08},
  journaltitle = {IEEE Transactions on Medical Imaging},
  volume = {21},
  number = {8},
  pages = {966--977},
  issn = {1558-254X},
  doi = {10.1109/TMI.2002.803106},
  url = {https://ieeexplore.ieee.org/document/1076041},
  urldate = {2025-07-22}
}

@article{Chen_ApplMagnReson_2017,
  title = {Joint {{Reconstruction}} of {{Multi-contrast Images}} and {{Multi-channel Coil Sensitivities}}},
  author = {Chen, Zhongzhou and Ren, Yanan and Su, Shi and Shi, Caiyun and Ji, Jim X. and Zheng, Hairong and Liu, Xin and Xie, Guoxi},
  date = {2017-09-01},
  journaltitle = {Appl Magn Reson},
  volume = {48},
  number = {9},
  pages = {955--969},
  issn = {1613-7507},
  doi = {10.1007/s00723-017-0919-4},
  url = {https://doi.org/10.1007/s00723-017-0919-4},
  urldate = {2025-11-02},
  langid = {english}
}

@article{Committee_Magn.Reson.Med._2024,
  title = {Recommended Implementation of Quantitative Susceptibility Mapping for Clinical Research in the Brain: {{A}} Consensus of the {{ISMRM}} Electro-Magnetic Tissue Properties Study Group},
  shorttitle = {Recommended Implementation of Quantitative Susceptibility Mapping for Clinical Research in the Brain},
  author = {Committee, QSM Consensus Organization and Bilgic, Berkin and Costagli, Mauro and Chan, Kwok-Shing and Duyn, Jeff and Langkammer, Christian and Lee, Jongho and Li, Xu and Liu, Chunlei and Marques, Jos\'e P. and Milovic, Carlos and Robinson, Simon Daniel and Schweser, Ferdinand and Shmueli, Karin and Spincemaille, Pascal and Straub, Sina and family=Zijl, given=Peter, prefix=van, useprefix=true and Wang, Yi and Group, ISMRM Electro-Magnetic Tissue Properties Study},
  date = {2024},
  journaltitle = {Magnetic Resonance in Medicine},
  volume = {91},
  number = {5},
  pages = {1834--1862},
  issn = {1522-2594},
  doi = {10.1002/mrm.30006},
  url = {https://onlinelibrary.wiley.com/doi/abs/10.1002/mrm.30006},
  urldate = {2025-07-22},
  langid = {english}
}

@article{Daeun_Magn.Reson.Med._2021,
  title = {Region-Optimized Virtual ({{ROVir}}) Coils: {{Localization}} and/or Suppression of Spatial Regions Using Sensor-Domain Beamforming},
  author = {Kim, Daeun and Cauley, Stephen F. and Nayak, Krishna S. and Leahy, Richard M. and Haldar, Justin P.},
  date = {2021},
  journaltitle = {Magn. Reson. Med.},
  volume = {86},
  number = {1},
  eprint = {https://onlinelibrary.wiley.com/doi/pdf/10.1002/mrm.28706},
  pages = {197--212},
  doi = {10.1002/mrm.28706},
  url = {https://onlinelibrary.wiley.com/doi/abs/10.1002/mrm.28706}
}

@article{Dong_NMRBiomed._2023,
  title = {Water/Fat Separation for Self-Navigated Diffusion-Weighted Multishot Echo-Planar Imaging},
  author = {Dong, Yiming and Riedel, Malte and Koolstra, Kirsten and family=Osch, given=Matthias J. P., prefix=van, useprefix=true and B\"ornert, Peter},
  date = {2023},
  journaltitle = {NMR in Biomedicine},
  volume = {36},
  number = {1},
  pages = {e4822},
  issn = {1099-1492},
  doi = {10.1002/nbm.4822},
  url = {https://onlinelibrary.wiley.com/doi/abs/10.1002/nbm.4822},
  urldate = {2025-11-02},
  langid = {english}
}

@article{Dubljevic_Magn.Reson.Med._2024,
  title = {Effect of {{MR}} Head Coil Geometry on Deep-Learning-Based {{MR}} Image Reconstruction},
  author = {Dubljevic, Natalia and Moore, Stephen and Lauzon, Michel Louis and Souza, Roberto and Frayne, Richard},
  date = {2024},
  journaltitle = {Magnetic Resonance in Medicine},
  volume = {92},
  number = {4},
  pages = {1404--1420},
  issn = {1522-2594},
  doi = {10.1002/mrm.30130},
  url = {https://onlinelibrary.wiley.com/doi/abs/10.1002/mrm.30130},
  urldate = {2025-11-02},
  langid = {english}
}

@article{Feizollah_Magn.Reson.Med._2025,
  title = {{{3D MERMAID}}: {{3D Multi-shot}} Enhanced Recovery Motion Artifact Insensitive Diffusion for Submillimeter, Multi-Shell, and {{SNR-efficient}} Diffusion Imaging},
  shorttitle = {{{3D MERMAID}}},
  author = {Feizollah, Sajjad and Tardif, Christine L.},
  date = {2025},
  journaltitle = {Magnetic Resonance in Medicine},
  volume = {93},
  number = {6},
  pages = {2311--2330},
  issn = {1522-2594},
  doi = {10.1002/mrm.30436},
  url = {https://onlinelibrary.wiley.com/doi/abs/10.1002/mrm.30436},
  urldate = {2025-11-02},
  langid = {english}
}

@article{FengZhao_IEEETrans.Med.Imaging_2012,
  title = {Separate {{Magnitude}} and {{Phase Regularization}} via {{Compressed Sensing}}},
  author = {{Feng Zhao} and Noll, D. C. and Nielsen, J-F and Fessler, J. A.},
  date = {2012-09},
  journaltitle = {IEEE Trans. Med. Imaging},
  volume = {31},
  number = {9},
  pages = {1713--1723},
  issn = {0278-0062, 1558-254X},
  doi = {10.1109/TMI.2012.2196707},
  url = {http://ieeexplore.ieee.org/document/6190747/},
  urldate = {2025-09-21}
}

@article{Griswold_Magn.Reson.Med._2002,
  title = {Generalized Autocalibrating Partially Parallel Acquisitions ({{GRAPPA}})},
  author = {Griswold, Mark A. and Jakob, Peter M. and Heidemann, Robin M. and Nittka, Mathias and Jellus, Vladimir and Wang, Jianmin and Kiefer, Berthold and Haase, Axel},
  date = {2002-06},
  journaltitle = {Magn Reson Med},
  volume = {47},
  number = {6},
  pages = {1202--1210},
  publisher = {Wiley Online Library},
  doi = {10.1002/mrm.10171},
  langid = {english}
}

@article{Griswold_Proc.Annu.Meet.ISMRM_2002,
  title = {The {{Use}} of an {{Adaptive Reconstruction}} for {{Array Coil Sensitivity Mapping}} and {{Intensity Normalization}}},
  author = {Griswold, Mark and Walsh, David and Heidemann, Robin and Haase, Axel and Jakob, Peter},
  date = {2002},
  journaltitle = {Proceedings of the Annual Meeting of ISMRM},
  volume = {10},
  pages = {2410},
  langid = {english}
}

@article{Haldar_IEEETrans.Med.Imaging_2014,
  title = {Low-Rank Modeling of Local k-{{Space}} Neighborhoods ({{LORAKS}}) for Constrained {{MRI}}},
  author = {Haldar, J P},
  date = {2014},
  journaltitle = {IEEE Trans. Med. Imag.},
  volume = {33},
  pages = {668--681},
  doi = {10.1109/TMI.2013.2293974}
}

@inproceedings{Holme_Proc.Annu.Meet.ISMRM_2023,
  title = {Non-{{Linear Reconstruction}} for {{Coil Sensitivity Calibration}} from {{Cartesian}} and Non-{{Cartesian Data}}},
  booktitle = {Proceedings of the {{Annual Meeting}} of {{ISMRM}}},
  author = {Holme, H Christian M and Uecker, Martin},
  date = {2023},
  pages = {4616},
  location = {Toronto}
}

@article{Holme_Sci.Rep._2019,
  title = {{{ENLIVE}}: {{An Efficient Nonlinear Method}} for {{Calibrationless}} and {{Robust Parallel Imaging}}},
  author = {Holme, H. Christian M. and Rosenzweig, Sebastian and Ong, Frank and Wilke, Robin N. and Lustig, Michael and Uecker, Martin},
  date = {2019-02-28},
  journaltitle = {Sci Rep},
  volume = {9},
  number = {1},
  pages = {3034},
  issn = {2045-2322},
  doi = {10.1038/s41598-019-39888-7},
  langid = {english},
  refid = {Holme2019}
}

@article{Hoult_J.Magn.Reson._1976a,
  title = {The Signal-to-Noise Ratio of the Nuclear Magnetic Resonance Experiment},
  author = {Hoult, D. I and Richards, R. E},
  date = {1976-10-01},
  journaltitle = {J. Magn. Reson.},
  volume = {24},
  number = {1},
  pages = {71--85},
  issn = {0022-2364},
  doi = {10.1016/0022-2364(76)90233-X},
  url = {https://www.sciencedirect.com/science/article/pii/002223647690233X},
  urldate = {2024-02-27}
}

@inproceedings{Inati_Proc.Annu.Meet.ISMRM_2013,
  title = {A {{Solution}} to the {{Phase Problem}} in {{Adaptive Coil Combination}}},
  booktitle = {Proceedings of the {{Annual Meeting}} of {{ISMRM}}},
  author = {Inati, Souheil J and Hansen, Michael S and Kellman, Peter},
  date = {2013},
  location = {Salt Lake City},
  url = {https://cds.ismrm.org/protected/13MProceedings/PDFfiles/2672.PDF},
  urldate = {2025-07-22}
}

@article{Isaieva_SciData_2021,
  title = {Multimodal Dataset of Real-Time {{2D}} and Static {{3D MRI}} of Healthy {{French}} Speakers},
  author = {Isaieva, Karyna and Laprie, Yves and Lecl\`ere, Justine and Douros, Ioannis K. and Felblinger, Jacques and Vuissoz, Pierre-Andr\'e},
  date = {2021-10-01},
  journaltitle = {Sci Data},
  volume = {8},
  number = {1},
  pages = {258},
  publisher = {Nature Publishing Group},
  issn = {2052-4463},
  doi = {10.1038/s41597-021-01041-3},
  url = {https://www.nature.com/articles/s41597-021-01041-3},
  urldate = {2025-11-02},
  langid = {english}
}

@article{Joseph_J.Magn.Reson.Imaging_2014,
  title = {Real-Time Flow {{MRI}} of the Aorta at a Resolution of 40 Msec},
  author = {Joseph, A. and Kowallick, J. T. and Merboldt, K.-D. and Voit, D. and Schaetz, S. and Zhang, S. and Sohns, J. M. and Lotz, J. and Frahm, J.},
  date = {2014},
  journaltitle = {J. Magn. Reson. Imaging},
  volume = {40},
  number = {1},
  pages = {206--213},
  issn = {1522-2586},
  doi = {10.1002/jmri.24328}
}

@article{Joseph_NMRBiomed._2011,
  title = {Real-Time Phase-Contrast {{MRI}} of Cardiovascular Blood Flow Using Undersampled Radial Fast Low-Angle Shot and Nonlinear Inverse Reconstruction},
  author = {Joseph, A. A. and Merboldt, K.-D. and Voit, D. and Zhang, S. and Uecker, M. and Lotz, J. and Frahm, J.},
  date = {2011},
  journaltitle = {NMR Biomed.},
  volume = {25},
  number = {7},
  pages = {917--924},
  publisher = {Wiley-Blackwell},
  issn = {0952-3480},
  doi = {10.1002/nbm.1812}
}

@article{Klimes_Magn.Reson.Med._2025,
  title = {Quantifying Spatial and Dynamic Lung Abnormalities with {{3D PREFUL FLORET UTE}} Imaging: {{A}} Feasibility Study},
  shorttitle = {Quantifying Spatial and Dynamic Lung Abnormalities with {{3D PREFUL FLORET UTE}} Imaging},
  author = {Klime\v s, Filip and Plummer, Joseph W. and Willmering, Matthew M. and Matheson, Alexander M. and Bdaiwi, Abdullah S. and Gutberlet, Marcel and Voskrebenzev, Andreas and Wernz, Marius M. and Wacker, Frank and Woods, Jason and Cleveland, Zackary I. and Walkup, Laura L. and Vogel-Claussen, Jens},
  date = {2025},
  journaltitle = {Magnetic Resonance in Medicine},
  volume = {93},
  number = {5},
  pages = {1984--1998},
  issn = {1522-2594},
  doi = {10.1002/mrm.30416},
  url = {https://onlinelibrary.wiley.com/doi/abs/10.1002/mrm.30416},
  urldate = {2025-11-02},
  langid = {english}
}

@article{Knoll_Magn.Reson.Med._2012,
  title = {Parallel Imaging with Nonlinear Reconstruction Using Variational Penalties},
  author = {Knoll, Florian and Clason, Christian and Bredies, Kristian and Uecker, Martin and Stollberger, Rudolf},
  date = {2012},
  journaltitle = {Magn Reson Med},
  volume = {67},
  number = {1},
  pages = {34--41},
  publisher = {Wiley Online Library},
  doi = {10.1002/mrm.22964},
  langid = {english}
}

@article{Laubrock_EuropeanJournalofRadiologyOpen_2022,
  title = {Imaging of Arrhythmia: {{Real-time}} Cardiac Magnetic Resonance Imaging in Atrial Fibrillation},
  shorttitle = {Imaging of Arrhythmia},
  author = {Laubrock, Kerstin and family=Loesch, given=Thassilo, prefix=von, useprefix=true and Steinmetz, Michael and Lotz, Joachim and Frahm, Jens and Uecker, Martin and Unterberg-Buchwald, Christina},
  date = {2022-01-01},
  journaltitle = {European Journal of Radiology Open},
  volume = {9},
  pages = {100404},
  issn = {2352-0477},
  doi = {10.1016/j.ejro.2022.100404},
  url = {https://www.sciencedirect.com/science/article/pii/S2352047722000119},
  urldate = {2025-11-02}
}

@article{Lee_Magn.Reson.Med._2024,
  title = {Replication of the {{bSTAR}} Sequence and Open-Source Implementation},
  author = {Lee, Nam G. and Bauman, Grzegorz and Bieri, Oliver and Nayak, Krishna S.},
  date = {2024},
  journaltitle = {Magnetic Resonance in Medicine},
  volume = {91},
  number = {4},
  pages = {1464--1477},
  issn = {1522-2594},
  doi = {10.1002/mrm.29947},
  url = {https://onlinelibrary.wiley.com/doi/abs/10.1002/mrm.29947},
  urldate = {2025-11-02},
  langid = {english}
}

@article{Leynes_IEEETrans.Med.Imaging_2024,
  title = {Scan-{{Specific Self-Supervised Bayesian Deep Non-Linear Inversion}} for {{Undersampled MRI Reconstruction}}},
  author = {Leynes, Andrew P. and Deveshwar, Nikhil and Nagarajan, Srikantan S. and Larson, Peder E. Z.},
  date = {2024-06},
  journaltitle = {IEEE Transactions on Medical Imaging},
  volume = {43},
  number = {6},
  pages = {2358--2369},
  issn = {1558-254X},
  doi = {10.1109/TMI.2024.3364911},
  url = {https://ieeexplore.ieee.org/abstract/document/10431784/references},
  urldate = {2025-11-02}
}

@article{Li_J.Magn.Reson.Imaging_2015,
  title = {Artifactual Microhemorrhage Generated by Susceptibility Weighted Image Processing},
  author = {Li, Ningzhi and Wang, Wen-Tung and Pham, Dzung L. and Butman, John A.},
  date = {2015},
  journaltitle = {J. Magn. Reson. Imaging},
  volume = {41},
  number = {6},
  pages = {1695--1700},
  issn = {1522-2586},
  doi = {10.1002/jmri.24728}
}

@article{Liu_BMEFront._2021,
  title = {Real-{{Time High-Resolution MRI Endoscopy}} at up to 10 {{Frames}} per {{Second}}},
  author = {Liu, Xiaoyang and Karmarkar, Parag and Voit, Dirk and Frahm, Jens and Weiss, Clifford R. and Kraitchman, Dara L. and Bottomley, Paul A.},
  date = {2021-02-25},
  journaltitle = {BME Frontiers},
  volume = {2021},
  pages = {6185616},
  publisher = {American Association for the Advancement of Science},
  doi = {10.34133/2021/6185616},
  url = {https://spj.science.org/doi/full/10.34133/2021/6185616},
  urldate = {2025-11-02}
}

@article{Liu_IEEETrans.Comput.Imaging_2021,
  title = {{{PALMNUT}}: {{An Enhanced Proximal Alternating Linearized Minimization Algorithm With Application}} to {{Separate Regularization}} of {{Magnitude}} and {{Phase}}},
  shorttitle = {{{PALMNUT}}},
  author = {Liu, Yunsong and Haldar, Justin P.},
  date = {2021},
  journaltitle = {IEEE Trans. Comput. Imaging},
  volume = {7},
  pages = {530--518},
  issn = {2333-9403, 2334-0118, 2573-0436},
  doi = {10.1109/TCI.2021.3077806},
  url = {https://ieeexplore.ieee.org/document/9424957/},
  urldate = {2025-09-21}
}

@article{Lobos_IEEETrans.Med.Imaging_2024,
  title = {New {{Theory}} and {{Faster Computations}} for {{Subspace-Based Sensitivity Map Estimation}} in {{Multichannel MRI}}},
  author = {Lobos, Rodrigo A. and Chan, Chin-Cheng and Haldar, Justin P.},
  date = {2024-01},
  journaltitle = {IEEE Transactions on Medical Imaging},
  volume = {43},
  number = {1},
  pages = {286--296},
  issn = {1558-254X},
  doi = {10.1109/TMI.2023.3297851},
  url = {https://ieeexplore.ieee.org/abstract/document/10190117},
  urldate = {2025-08-06}
}

@article{Lustig_Magn.Reson.Med._2007,
  title = {Sparse {{MRI}}: {{The}} Application of Compressed Sensing for Rapid {{MR}} Imaging},
  author = {Lustig, M. and Donoho, D. and Pauly, J. M.},
  date = {2007},
  journaltitle = {Magn. Reson. Med.},
  volume = {58},
  number = {6},
  pages = {1182--1195},
  publisher = {Wiley Online Library}
}

@article{Lustig_Magn.Reson.Med._2010,
  title = {{{SPIRiT}}: {{Iterative}} Self-Consistent Parallel Imaging Reconstruction from Arbitrary k-Space},
  author = {Lustig, M. and Pauly, J. M.},
  date = {2010},
  journaltitle = {Magn. Reson. Med.},
  volume = {64},
  number = {2},
  pages = {457--471}
}

@article{Maier_Magn.Reson.Med._2018,
  title = {Rapid {{T1}} Quantification from High Resolution {{3D}} Data with Model-Based Reconstruction},
  author = {Maier, Oliver and Schoormans, Jasper and Schloegl, Matthias and Strijkers, Gustav J. and Lesch, Andreas and Benkert, Thomas and Block, Tobias and Coolen, Bram F. and Bredies, Kristian and Stollberger, Rudolf},
  date = {2019},
  journaltitle = {Magn. Reson. Med.},
  volume = {81},
  number = {3},
  pages = {2072--2089},
  doi = {10.1002/mrm.27502}
}

@article{Niebergall_Magn.Reson.Med._2013,
  title = {Real-Time {{MRI}} of Speaking at a Resolution of 33 Ms: {{Undersampled}} Radial {{FLASH}} with Nonlinear Inverse Reconstruction},
  author = {Niebergall, A. and Zhang, S. and Kunay, E. and Keydana, G. and Job, M. and Uecker, M. and Frahm, J.},
  date = {2013},
  journaltitle = {Magn. Reson. Med.},
  volume = {69},
  number = {2},
  pages = {477--485},
  publisher = {Wiley Online Library}
}

@article{Olthoff_GastroenterologyResearchandPractice_2014,
  title = {On the {{Physiology}} of {{Normal Swallowing}} as {{Revealed}} by {{Magnetic Resonance Imaging}} in {{Real Time}}},
  author = {Olthoff, A. and Zhang, S. and Schweizer, R. and Frahm, J.},
  date = {2014},
  journaltitle = {Gastroenterology Research and Practice},
  volume = {2014},
  pages = {10},
  doi = {10.1155/2014/493174}
}

@article{Ong_Magn.Reson.Med._2017,
  title = {General Phase Regularized Reconstruction Using Phase Cycling},
  author = {Ong, Frank and Cheng, Joseph Y. and Lustig, Michael},
  date = {2017},
  journaltitle = {Magn. Reson. Med.},
  issn = {1522-2594},
  doi = {10.1002/mrm.27011}
}

@article{Parker_Magn.Reson.Med._2014,
  title = {Phase Reconstruction from Multiple Coil Data Using a Virtual Reference Coil},
  author = {Parker, Dennis L. and Payne, Allison and Todd, Nick and Hadley, J. Rock},
  date = {2014},
  journaltitle = {Magnetic Resonance in Medicine},
  volume = {72},
  number = {2},
  pages = {563--569},
  issn = {1522-2594},
  doi = {10.1002/mrm.24932},
  url = {https://onlinelibrary.wiley.com/doi/abs/10.1002/mrm.24932},
  urldate = {2025-07-22},
  langid = {english}
}

@article{Pruessmann_Magn.Reson.Med._1999,
  title = {{{SENSE}}: Sensitivity Encoding for Fast {{MRI}}},
  author = {Pruessmann, K. P. and Weiger, M. and Scheidegger, M. B. and Boesiger, P.},
  date = {1999},
  journaltitle = {Magn. Reson. Med.},
  volume = {42},
  number = {5},
  pages = {952--962}
}

@article{Pruessmann_Magn.Reson.Med._2001,
  title = {Advances in Sensitivity Encoding with Arbitrary K-Space Trajectories},
  author = {Pruessmann, K. P. and Weiger, M. and Boernert, P. and Boesiger, P.},
  date = {2001},
  journaltitle = {Magn. Reson. Med.},
  volume = {46},
  number = {4},
  pages = {638--651},
  publisher = {Wiley Online Library}
}

@article{Robinson_NMRBiomed._2017,
  title = {An Illustrated Comparison of Processing Methods for {{MR}} Phase Imaging and {{QSM}}: Combining Array Coil Signals and Phase Unwrapping},
  shorttitle = {An Illustrated Comparison of Processing Methods for {{MR}} Phase Imaging and {{QSM}}},
  author = {Robinson, Simon Daniel and Bredies, Kristian and Khabipova, Diana and Dymerska, Barbara and Marques, Jos\'e P. and Schweser, Ferdinand},
  date = {2017},
  journaltitle = {NMR in Biomedicine},
  volume = {30},
  number = {4},
  pages = {e3601},
  issn = {1099-1492},
  doi = {10.1002/nbm.3601},
  url = {https://onlinelibrary.wiley.com/doi/abs/10.1002/nbm.3601},
  urldate = {2025-07-22},
  langid = {english}
}

@article{Roemer_Magn.Reson.Med._1990,
  title = {The {{NMR}} Phased Array},
  author = {Roemer, P. B. and Edelstein, W. A. and Hayes, C. E. and Souza, S. P. and Mueller, O. M.},
  date = {1990-11},
  journaltitle = {Magn. Reson. Med.},
  volume = {16},
  number = {2},
  pages = {192--225},
  publisher = {Wiley Online Library},
  doi = {10.1002/mrm.1910160203},
  langid = {english}
}

@article{Rosenzweig_Magn.Reson.Med._2018,
  title = {Simultaneous Multi-Slice {{MRI}} Using Cartesian and Radial {{FLASH}} and Regularized Nonlinear Inversion: {{SMS-NLINV}}},
  author = {Rosenzweig, Sebastian and Holme, Hans Christian Martin and Wilke, Robin N. and Voit, Dirk and Frahm, Jens and Uecker, Martin},
  date = {2018},
  journaltitle = {Magn. Reson. Med.},
  volume = {79},
  number = {4},
  pages = {2057--2066},
  doi = {10.1002/mrm.26878}
}

@article{Rosenzweig_Magn.Reson.Med._2018a,
  title = {Simple Auto-calibrated Gradient Delay Estimation from Few Spokes Using {{Radial Intersections}} ({{RING}})},
  author = {Rosenzweig, Sebastian and Holme, H. Christian M. and Uecker, Martin},
  date = {2019},
  journaltitle = {Magn. Reson. Med.},
  volume = {81},
  number = {3},
  eprint = {https://onlinelibrary.wiley.com/doi/pdf/10.1002/mrm.27506},
  pages = {1898--1906},
  doi = {10.1002/mrm.27506},
  url = {https://onlinelibrary.wiley.com/doi/abs/10.1002/mrm.27506},
  langid = {english}
}

@article{Schaetz_ComputMathMethodM_2017,
  title = {Accelerated Computing in Magnetic Resonance Imaging: {{Real-time}} Imaging Using Nonlinear Inverse Reconstruction},
  author = {Schaetz, Sebastian and Voit, Dirk and Frahm, Jens and Uecker, Martin},
  date = {2017},
  journaltitle = {Comput. Math. Method Med.},
  volume = {2017},
  publisher = {Hindawi},
  doi = {10.1155/2017/3527269}
}

@article{Schaten_MagnResonMaterPhy_2024,
  title = {{{BART}} Streams: A Plug \& Play Framework for Interactive Real-Time {{MRI}} at the Example of Aligned Dynamic Coil Compression},
  author = {Schaten, Philip and Blumenthal, Moritz and Rapp, Bernhard and Uecker, Martin},
  date = {2024},
  journaltitle = {Magn Reson Mater Phy},
  volume = {37},
  number = {S1},
  pages = {719--721},
  issn = {1352-8661},
  doi = {10.1007/s10334-024-01191-6},
  url = {https://link.springer.com/10.1007/s10334-024-01191-6},
  urldate = {2025-08-03},
  langid = {english}
}

@article{Shin_Magn.Reson.Med._2014,
  title = {Calibrationless Parallel Imaging Reconstruction Based on Structured Low-Rank Matrix Completion},
  author = {Shin, Peter J. and Larson, Peder E. Z. and Ohliger, Michael A. and Elad, Michael and Pauly, John M. and Vigneron, Daniel B. and Lustig, Michael},
  date = {2014},
  journaltitle = {Magn. Reson. Med.},
  volume = {72},
  number = {4},
  pages = {959--970},
  issn = {1522-2594},
  doi = {10.1002/mrm.24997}
}

@article{Tasdelen_Magn.Reson.Med._2024,
  title = {Improved Abdominal {{T1}} Weighted Imaging at 0.{{55T}}},
  author = {Tasdelen, Bilal and Lee, Nam G. and Cui, Sophia X. and Nayak, Krishna S.},
  date = {2024},
  journaltitle = {Magnetic Resonance in Medicine},
  volume = {92},
  number = {6},
  pages = {2580--2587},
  issn = {1522-2594},
  doi = {10.1002/mrm.30224},
  url = {https://onlinelibrary.wiley.com/doi/abs/10.1002/mrm.30224},
  urldate = {2025-11-02},
  langid = {english}
}

@inproceedings{Uecker__2013,
  title = {Software Toolbox and Programming Library for Compressed Sensing and Parallel Imaging},
  booktitle = {{{ISMRM}} Workshop on Data Sampling and Image Reconstruction},
  author = {Uecker, M. and Virtue, P. and Ong, F. and Murphy, M. J. and Alley, M. T. and Vasanawala, S. S. and Lustig, M.},
  date = {2013},
  location = {Sedona}
}

@inproceedings{Uecker__2015,
  title = {Berkeley Advanced Reconstruction Toolbox},
  booktitle = {Proc. {{Intl}}. {{Soc}}. {{Mag}}. {{Reson}}. {{Med}}.},
  author = {Uecker, M. and Ong, F. and Tamir, J. I. and Bahri, D. and Virtue, P. and Cheng, J. Y. and Zhang, T. and Lustig, M.},
  date = {2015},
  volume = {23},
  pages = {2486},
  location = {Toronto}
}

@article{Uecker_Magn.Reson.Med._2008,
  title = {Image Reconstruction by Regularized Nonlinear Inversion-Joint Estimation of Coil Sensitivities and Image Content},
  author = {Uecker, M. and Hohage, T. and Block, K. T. and Frahm, J.},
  date = {2008},
  journaltitle = {Magn. Reson. Med.},
  volume = {60},
  number = {3},
  pages = {674--682},
  publisher = {Wiley Online Library}
}

@article{Uecker_Magn.Reson.Med._2010,
  title = {Nonlinear Inverse Reconstruction for Real-Time {{MRI}} of the Human Heart Using Undersampled Radial {{FLASH}}},
  author = {Uecker, M. and Zhang, S. and Frahm, J.},
  date = {2010},
  journaltitle = {Magn. Reson. Med.},
  volume = {63},
  number = {6},
  pages = {1456--1462},
  publisher = {Wiley-Blackwell},
  issn = {1522-2594},
  doi = {10.1002/mrm.22453}
}

@article{Uecker_Magn.Reson.Med._2014,
  title = {{{ESPIRiT}}---an Eigenvalue Approach to Autocalibrating Parallel {{MRI}}: {{Where SENSE}} Meets {{GRAPPA}}},
  author = {Uecker, Martin and Lai, Peng and Murphy, Mark J. and Virtue, Patrick and Elad, Michael and Pauly, John M. and Vasanawala, Shreyas S. and Lustig, Michael},
  date = {2014},
  journaltitle = {Magn. Reson. Med.},
  volume = {71},
  number = {3},
  pages = {990--1001},
  publisher = {Wiley Online Library},
  doi = {10.1002/mrm.24751},
  langid = {english}
}

@article{Uecker_Magn.Reson.Med._2017,
  title = {Estimating Absolute-Phase Maps Using {{ESPIRiT}} and Virtual Conjugate Coils},
  author = {Uecker, M. and Lustig, M.},
  date = {2017},
  journaltitle = {Magn. Reson. Med.},
  volume = {77},
  number = {3},
  pages = {1201--1207},
  issn = {1522-2594},
  doi = {10.1002/mrm.26191}
}

@article{Uecker_NMRBiomed._2010,
  title = {Real-Time {{MRI}} at a Resolution of 20 Ms},
  author = {Uecker, M. and Zhang, S. and Voit, D. and Karaus, A. and Merboldt, K.-D. and Frahm, J.},
  date = {2010},
  journaltitle = {NMR Biomed.},
  volume = {23},
  number = {8},
  pages = {986--994},
  publisher = {John Wiley \& Sons, Ltd.},
  issn = {1099-1492},
  doi = {10.1002/nbm.1585}
}

@article{Untenberger_Magn.Reson.Med._2016,
  title = {Advances in Real-Time Phase-Contrast Flow {{MRI}} Using Asymmetric Radial Gradient Echoes},
  author = {Untenberger, M. and Tan, Z. and Voit, D. and Joseph, A. A. and Roeloffs, V. and Merboldt, K. D. and Sch\"atz, S. and Frahm, J.},
  date = {2016},
  journaltitle = {Magn. Reson. Med.},
  volume = {75},
  number = {5},
  pages = {1901--1908},
  issn = {1522-2594},
  doi = {10.1002/mrm.25696}
}

@article{Voit_Quant.ImagingMed.Surg._2019,
  title = {Body Coil Reference for Inverse Reconstructions of Multi-Coil Data---the Case for Real-Time {{MRI}}},
  author = {Voit, Dirk and Kalentev, Oleksandr and Frahm, Jens},
  date = {2019-11},
  journaltitle = {Quant. Imaging Med. Surg.},
  volume = {9},
  number = {11},
  pages = {1815--1819},
  issn = {22234292, 22234306},
  doi = {10.21037/qims.2019.08.14},
  url = {http://qims.amegroups.com/article/view/29413/26547},
  urldate = {2025-08-03}
}

@article{Vu_Magn.Reson.Med._,
  title = {{{DREAMER}}: {{Rapid}} and {{Simultaneous Multiple Contrast Magnetic Resonance Imaging}} of {{Solid}} and {{Soft Tissue}}},
  shorttitle = {{{DREAMER}}},
  author = {Vu, Brian-Tinh Duc and Kamona, Nada and Wehrli, Felix W. and Deshpande, Rajiv S. and Song, Hee Kwon and Hu, Allison and Baccaglini, Emily and Bartlett, Scott P. and Rajapakse, Chamith S.},
  journaltitle = {Magnetic Resonance in Medicine},
  volume = {n/a},
  number = {n/a},
  issn = {1522-2594},
  doi = {10.1002/mrm.70140},
  url = {https://onlinelibrary.wiley.com/doi/abs/10.1002/mrm.70140},
  urldate = {2025-11-02},
  langid = {english}
}

@article{Vu_Magn.Reson.Med._2025,
  title = {Three Contrasts in 3 Min: {{Rapid}}, High-Resolution, and Bone-Selective {{UTE MRI}} for Craniofacial Imaging with Automated Deep-Learning Skull Segmentation},
  shorttitle = {Three Contrasts in 3 Min},
  author = {Vu, Brian-Tinh Duc and Kamona, Nada and Kim, Yohan and Ng, Jinggang J. and Jones, Brandon C. and Wehrli, Felix W. and Song, Hee Kwon and Bartlett, Scott P. and Lee, Hyunyeol and Rajapakse, Chamith S.},
  date = {2025},
  journaltitle = {Magnetic Resonance in Medicine},
  volume = {93},
  number = {1},
  pages = {245--260},
  issn = {1522-2594},
  doi = {10.1002/mrm.30275},
  url = {https://onlinelibrary.wiley.com/doi/abs/10.1002/mrm.30275},
  urldate = {2025-11-02},
  langid = {english}
}

@article{Walsh_Magn.Reson.Med._2000,
  title = {Adaptive Reconstruction of Phased Array {{MR}} Imagery},
  author = {Walsh, David O. and Gmitro, Arthur F. and Marcellin, Michael W.},
  date = {2000-05},
  journaltitle = {Magn. Reson. Med.},
  volume = {43},
  number = {5},
  pages = {682--690},
  publisher = {Wiley Online Library},
  doi = {10.1002/(SICI)1522-2594(200005)43:5<682::AID-MRM10>3.0.CO;2-G},
  langid = {english}
}

@article{Wang_Magn.Reson.Med._2018,
  title = {Model-Based {{T1}} Mapping with Sparsity Constraints Using Single-Shot Inversion-Recovery Radial {{FLASH}}},
  author = {Wang, Xiaoqing and Roeloffs, Volkert and Klosowski, Jakob and Tan, Zhengguo and Voit, Dirk and Uecker, Martin and Frahm, Jens},
  date = {2018},
  journaltitle = {Magn. Reson. Med.},
  volume = {79},
  number = {2},
  pages = {730--740},
  issn = {1522-2594},
  doi = {10.1002/mrm.26726}
}

@article{Wang_Magn.Reson.Med._2025,
  title = {Rapid, High-Resolution and Distortion-Free {{R2}}{$\ast$} Mapping of Fetal Brain Using Multi-Echo Radial {{FLASH}} and Model-Based Reconstruction},
  author = {Wang, Xiaoqing and Fan, Hongli and Tan, Zhengguo and Vasylechko, Serge and Yang, Edward and Didier, Ryne and Afacan, Onur and Uecker, Martin and Warfield, Simon K. and Gholipour, Ali},
  date = {2025},
  journaltitle = {Magnetic Resonance in Medicine},
  volume = {94},
  number = {5},
  pages = {1913--1929},
  issn = {1522-2594},
  doi = {10.1002/mrm.30604},
  url = {https://onlinelibrary.wiley.com/doi/abs/10.1002/mrm.30604},
  urldate = {2025-11-02},
  langid = {english}
}

@article{Wang_TechnolCancerResTreat_2017,
  title = {Accelerated {{Brain DCE-MRI Using Iterative Reconstruction With Total Generalized Variation Penalty}} for {{Quantitative Pharmacokinetic Analysis}}: {{A Feasibility Study}}},
  shorttitle = {Accelerated {{Brain DCE-MRI Using Iterative Reconstruction With Total Generalized Variation Penalty}} for {{Quantitative Pharmacokinetic Analysis}}},
  author = {Wang, Chunhao and Yin, Fang-Fang and Kirkpatrick, John P. and Chang, Zheng},
  date = {2017-08-01},
  journaltitle = {Technol Cancer Res Treat},
  volume = {16},
  number = {4},
  pages = {446--460},
  publisher = {SAGE Publications Inc},
  issn = {1533-0346},
  doi = {10.1177/1533034616649294},
  url = {https://doi.org/10.1177/1533034616649294},
  urldate = {2025-11-02},
  langid = {english}
}

@article{Xiang_J.Magn.Reson.Imaging_2023,
  title = {Balanced {{Steady-State Free Precession Cine MR Imaging}} in the {{Presence}} of {{Cardiac Devices}}: {{Value}} of {{Interleaved Radial Linear Combination Acquisition With Partial Dephasing}}},
  shorttitle = {Balanced {{Steady-State Free Precession Cine MR Imaging}} in the {{Presence}} of {{Cardiac Devices}}},
  author = {Xiang, Jie and Lamy, Jerome and Lampert, Rachel and Peters, Dana C.},
  date = {2023},
  journaltitle = {Journal of Magnetic Resonance Imaging},
  volume = {58},
  number = {3},
  pages = {782--791},
  issn = {1522-2586},
  doi = {10.1002/jmri.28528},
  url = {https://onlinelibrary.wiley.com/doi/abs/10.1002/jmri.28528},
  urldate = {2025-11-02},
  langid = {english}
}

@article{Xu_Magn.Reson.Med._2013,
  title = {Fast {{3D}} Contrast Enhanced {{MRI}} of the Liver Using Temporal Resolution Acceleration with Constrained Evolution Reconstruction},
  author = {Xu, Bo and Spincemaille, Pascal and Chen, Gang and Agrawal, Mukta and Nguyen, Thanh D. and Prince, Martin R. and Wang, Yi},
  date = {2013},
  journaltitle = {Magnetic Resonance in Medicine},
  volume = {69},
  number = {2},
  pages = {370--381},
  issn = {1522-2594},
  doi = {10.1002/mrm.24253},
  url = {https://onlinelibrary.wiley.com/doi/abs/10.1002/mrm.24253},
  urldate = {2025-11-02},
  langid = {english}
}

@article{Ying_Magn.Reson.Med._2007,
  title = {Joint Image Reconstruction and Sensitivity Estimation in {{SENSE}} ({{JSENSE}})},
  author = {Ying, Leslie and Sheng, Jinhua},
  date = {2007-05-29},
  journaltitle = {Magn. Reson. Med.},
  volume = {57},
  number = {6},
  pages = {1196--1202},
  publisher = {Wiley Online Library},
  doi = {10.1002/mrm.21245},
  langid = {english}
}

@article{Zeng_SignalProcessing:ImageCommunication_2020,
  title = {A Comparative Study of {{CNN-based}} Super-Resolution Methods in {{MRI}} Reconstruction and Its Beyond},
  author = {Zeng, Wei and Peng, Jie and Wang, Shanshan and Liu, Qiegen},
  date = {2020-02-01},
  journaltitle = {Signal Processing: Image Communication},
  volume = {81},
  pages = {115701},
  issn = {0923-5965},
  doi = {10.1016/j.image.2019.115701},
  url = {https://www.sciencedirect.com/science/article/pii/S0923596519302358},
  urldate = {2025-11-02}
}

@article{Zhang_J.Magn.Reson.Imaging_2012,
  title = {Real-Time Magnetic Resonance Imaging of Normal Swallowing},
  author = {Zhang, S. and Olthoff, A. and Frahm, J.},
  date = {2012},
  journaltitle = {J. Magn. Reson. Imaging},
  volume = {35},
  number = {6},
  pages = {1372--1379},
  publisher = {Wiley Subscription Services, Inc., A Wiley Company},
  issn = {1522-2586},
  doi = {10.1002/jmri.23591}
}

@article{Zhang_Magn.Reson.Med._2013,
  title = {Coil Compression for Accelerated Imaging with {{Cartesian}} Sampling},
  author = {Zhang, Tao and Pauly, John M. and Vasanawala, Shreyas S. and Lustig, Michael},
  date = {2013},
  journaltitle = {Magn. Reson. Med.},
  volume = {69},
  number = {2},
  pages = {571--582},
  publisher = {Wiley Subscription Services, Inc., A Wiley Company},
  issn = {1522-2594},
  doi = {10.1002/mrm.24267}
}

\clearpage
\appendix
\section*{Supporting Information}

\subsection*{Video S1}
Real-time reconstruction of the dataset shown in \figref{fig:figure_realtime} without (left) and with (right) phase pole correction.

\section*{Supporting Figures}

\begin{figure}[h!]
\includegraphics[width=\linewidth]{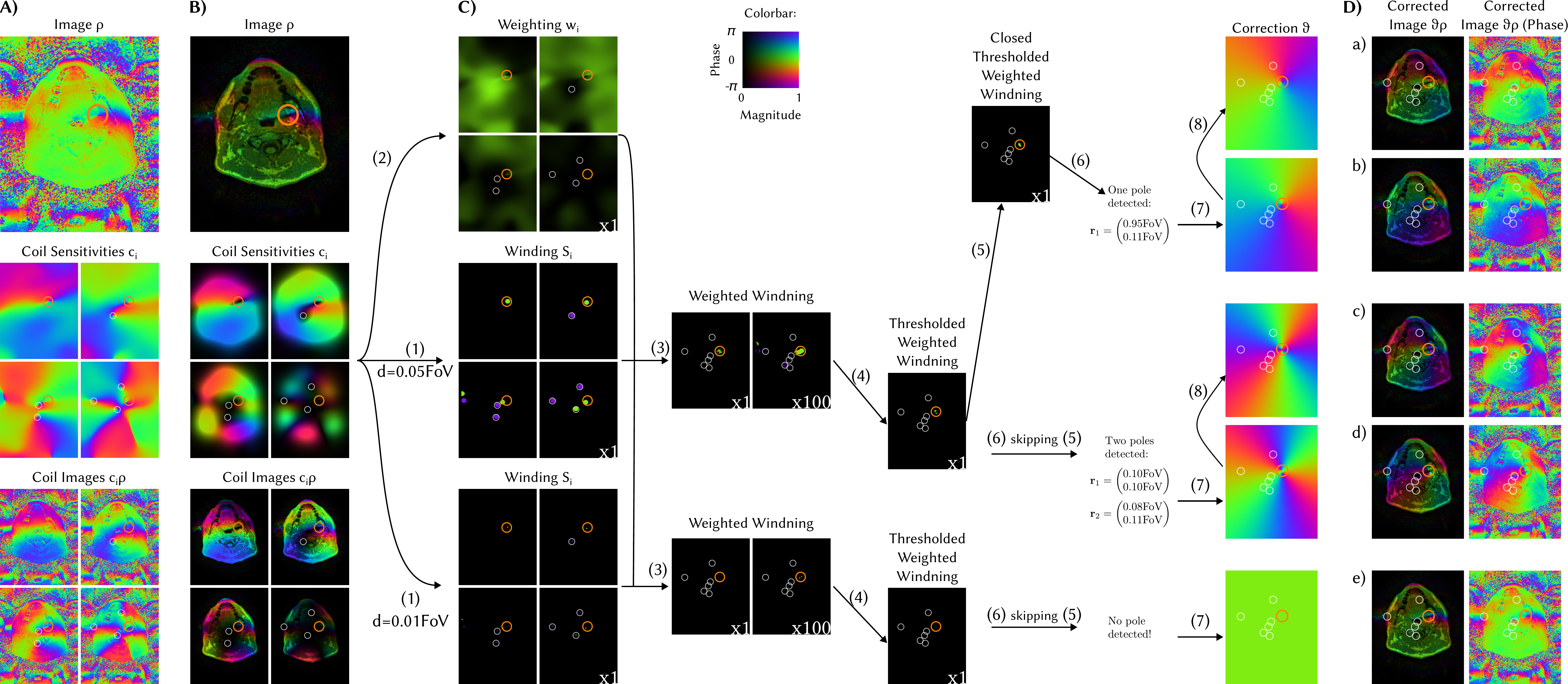}
\caption{Phase pole detection at the example of a real dataset. The image, coil sensitivities and their product
of the NLINV reconstruction in \figref{fig:figure_iterations} after 8 Gauss-Newton steps are shown in phase (A) and magnitude + phase representation (B).
The orange circles mark the position of the phase pole in the image, white circles mark the position of phase poles in the coil images.
(C) illustrates the steps of the detection and (D) shows the final corrected images. The final images show the effect of changing specific steps of the algorithm.
Namely, in (De) the circle diameter is reduced such that the poles detected in the respective coils do not overlap leading to no detected pole after the thresholding.
In (Dc) and (Dd), the closing operation is skipped leading to two distinct detected poles. The final correction over-compensates the pole present in the image leading to an opposite pole.
For (Db) and (Dd), the global phase selection is skipped, leading to a generally larger change of the overall image phase. (Da) shows the final corrected image with all steps as presented in
\figref{fig:figure_iterations}.}
\label{supfig:detection_real}
\end{figure}

\begin{figure}[h!]
\centering \includegraphics[width=\linewidth]{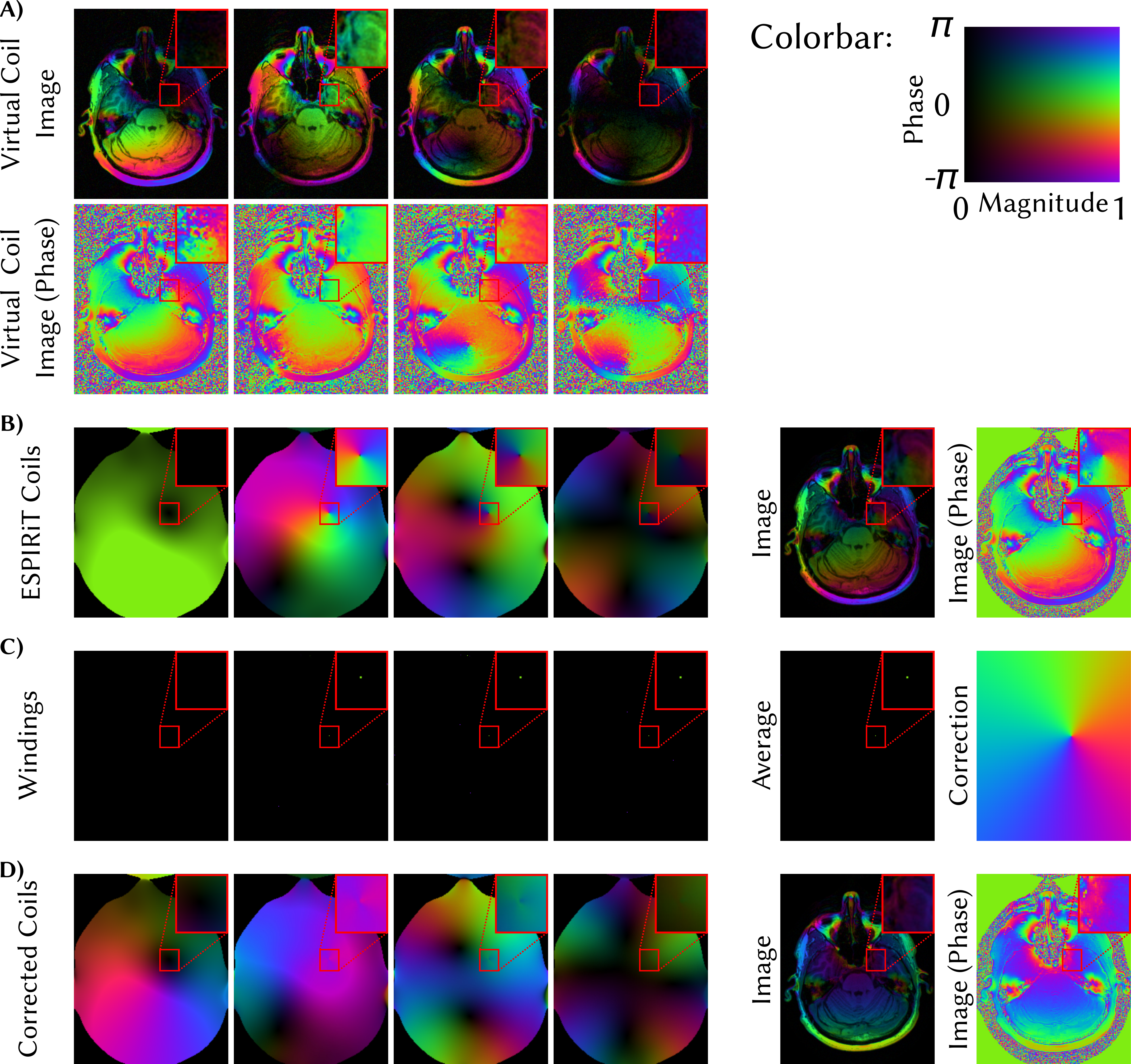}
\caption{Application of the phase pole correction to ESPIRiT coils.
(A) The first four coil images after coil compression. The first virtual coil image has a phase pole in the red square.
(B) ESPIRiT coil sensitivities and coil-combined image using these ESPIRiT coils.
As the first virtual coil is used as reference (it has phase zero), the phase pole is also present in the coil combined image which has the phase of the first coil image.
An opposite phase pole is visible in all other coil sensitivities to compensate the pole in the image.
(C) Phase pole correction is performed on the coil sensitivities in (B). The diameter to compute the winding number is set to 1 pixel as the phase pole is for ESPIRiT coils at the exactly same position in all coils.
(D) Corrected coils and corresponding coil combined image. The phase pole is removed, however, the corrected phase in the coils is not perfectly smooth.
}
\label{supfig:espirit}
\end{figure}


\end{document}